\documentclass[preprint,3p,times]{elsarticle}
\usepackage{wrapfig}
\usepackage{graphicx}
\usepackage{subcaption}

\usepackage{amssymb}

\usepackage[figuresright]{rotating}
 
\usepackage{tikz}
\usetikzlibrary{arrows.meta}

\usepackage{eucal}
\usepackage[utf8]{inputenc}
\usepackage{amsmath}

\usepackage{xcolor,colortbl}
\usepackage{caption}
\usepackage{graphicx}
\usepackage{natbib}
\usepackage{amsmath}
\usepackage{amssymb}
\usepackage{latexsym}
\usepackage{wasysym}
\usepackage{fontenc}
\usepackage{rotating}
\usepackage{array}

\newcommand{\narsimha}[1]{{\color{black}{#1}}}

\newsavebox{\astrutbox}
\sbox{\astrutbox}{\rule[-5pt]{0pt}{20pt}}

\newcommand{\be}{\begin{equation}}
\newcommand{\ee}{\end{equation}}
\newcommand{\bee}{\begin{eqnarray}}
\newcommand{\eee}{\end{eqnarray}}

\newcommand {\delux}[2]{\frac{\partial{#1}}{\partial{#2}}}
\newcommand {\bdelux}[2]{\dfrac{\partial{#1}}{\partial{#2}}}

\newcommand {\delsux}[2]{\frac{\partial^2{#1}}{\partial{{#2}^2}}}

\newcommand {\ui}{u_i}

\newcommand {\sgn}{\texttt{sign}}

\newcommand {\rs}{\rho^*}

\newcommand {\rb}{\overline{\rho}}

\definecolor{Gray}{gray}{0.85}
\definecolor{LightCyan}{rgb}{0.88,1,1}
\definecolor{Red}{rgb}{1,.5,0}
\newcolumntype{g}{>{\columncolor{Gray}}c}
\newcolumntype{y}{>{\columncolor{LightCyan}}c}
\newcolumntype{o}{>{\columncolor{Red}}c}

\begin{document}

\begin{frontmatter}



%
%

\title{Stable, Compact, and Direct Ghost-Cell Reconstruction: A Non-Iterative Approach for Embedded-Boundary Methods}
\author[label_math,label_entc]{Narsimha Reddy Rapaka\corref{cor1}}
\author[label_mae,label_entc]{Pankaj Jagad}
\author[label_mae,label_entc]{Yacine Addad}
\author[label_math,label_mci,label_entc]{Mohamed Kamel Riahi}
 
\cortext[cor1]{Corresponding author. Tel.: +971 2 3125416, 
\textit{Email address:} narsimha.rapaka@ku.ac.ae (Narsimha R. Rapaka)
}
\address[label_math]{Department of Mathematics, College of Computing and Mathematical Sciences, Khalifa University of Science and Technology, Abu Dhabi, UAE.}
\address[label_mae]{Department of Mechanical and Nuclear Engineering, Khalifa University of Science and Technology, Abu Dhabi, UAE.}
\address[label_mci]{Research Center for Mathematical \& Computational Intelligence (MCI), Khalifa University of Science and Technology, Abu Dhabi, UAE.}
\address[label_entc]{Emirates Nuclear Technology Center,  Khalifa University of Science and Technology, Abu Dhabi, UAE.}

\begin{abstract}
A direct, analytical, and non-iterative ghost-cell reconstruction framework
is developed for Cartesian-grid embedded-boundary methods. Analytical
expressions are derived for imposing Dirichlet and Neumann boundary conditions
in two and three dimensions directly at the embedded boundary. The formulation
eliminates the intermediate image-point reconstruction, matrix inversion, and
the need to precompute and store geometry-dependent reconstruction weights.
For the Cartesian reconstruction stencil considered here, the dependencies
among neighboring ghost cells form a directed acyclic graph. A topological
ordering partitions the ghost cells into dependency levels, enabling
level-by-level reconstruction without iterative ghost-cell updates. The
formulation is combined with hybrid ghost cells (HGC), whose centers may lie
on either side of the embedded boundary. This placement allows the
reconstruction to satisfy the linear reconstruction-stability criterion
previously derived for scalar advection while retaining a compact
nearest-neighbour Cartesian stencil. An extended-stencil classical ghost-cell
formulation (CGC\_ES) is included as a stability-preserving reference,
allowing the effects of reconstruction stability and stencil compactness to
be distinguished.

The same analytical reconstruction relations provide solution values and
spatial gradients directly on the embedded boundary, enabling the evaluation
of pressure forces, wall stresses, drag, and lift without a separate surface
reconstruction or filtering procedure. Simulations of flow past a circular
cylinder and an airfoil show that this linear stability criterion remains a
useful indicator of reconstruction stability for the nonlinear incompressible
Navier--Stokes cases considered. Classical ghost-cell reconstruction develops
spurious oscillations when the criterion is violated, whereas CGC\_ES and HGC
produce stable solutions. HGC provides the additional advantage of satisfying
the same stability requirement while retaining the compact nearest-neighbour
stencil.
\end{abstract}


\end{frontmatter}


\section{Introduction}

The immersed boundary method (IBM) provides an attractive framework for
simulating flows around complex geometries on Cartesian grids, avoiding many
of the mesh-generation difficulties associated with body-fitted
discretizations. Originally introduced by Peskin~\cite{Peskin:1972}, immersed
and embedded boundary methods have subsequently been applied to a wide range
of fluid-flow problems involving complex stationary and moving interfaces
\cite{Mittal_AR:2005,Sotiropoulos:2014}. In discrete embedded-boundary
formulations, boundary conditions are commonly imposed through reconstruction
at ghost cells located near the interface. The accuracy and robustness of the
method therefore depend not only on the underlying flow discretization but
also on the geometry, stability, and conditioning of the ghost-cell
reconstruction.

The location of a ghost cell relative to the embedded boundary determines the
reconstruction geometry and the associated interpolation weights. Moreover,
the reconstruction stencil of one ghost cell may contain neighboring ghost
cells whose values are themselves unknown. These features can lead to
geometry-dependent reconstruction stencils, storage of interpolation weights,
coupling among ghost cells, and, in some implementations, iterative
reconstruction procedures. The reconstruction geometry can also affect the
stability of the overall discrete system, particularly when the embedded
boundary lies unfavorably with respect to the Cartesian stencil
\cite{RapakaS:2018}. Consequently, ghost-cell placement, stencil compactness,
reconstruction stability, and inter-ghost-cell dependencies should be
considered together when designing an embedded-boundary reconstruction
procedure.

A widely used ghost-cell approach is the image-point method
\cite{Balaras:2004,MittalDBNVL:2008}. In this formulation, ghost cells are
located on the solid side of the embedded boundary and are mirrored across the
interface to define corresponding image points in the fluid domain. The
solution is first reconstructed at the image point and is subsequently used,
together with the boundary condition, to determine the ghost-cell value.
Mittal et al.~\cite{MittalDBNVL:2008}, for example, employed such an
image-point formulation together with a coupled iterative reconstruction
procedure. Tsai et al.~\cite{Tsai:2020} similarly used inverse-distance
weights and an iterative procedure to reconstruct the solution first at an
image point and subsequently at the ghost cell. Although image-point methods
provide considerable geometrical flexibility, the image-point location and
its interpolation stencil vary with the local interface geometry, and
neighboring ghost cells may enter the reconstruction procedure.

In the present work, we exploit the regular geometry of the Cartesian grid to
derive analytical expressions for direct ghost-cell reconstruction. Unlike
image-point approaches, the boundary condition is imposed directly at the
intersection of the embedded boundary with the boundary normal passing
through the ghost cell, thereby eliminating the intermediate image-point
reconstruction. The resulting formulation avoids matrix inversion,
pre-computed interpolation weights, and storage of reconstruction
coefficients.

The dependency structure generated by the Cartesian reconstruction stencil is
also exploited. For the stencil considered here, the ghost-cell dependencies
form a directed acyclic graph, permitting a topological ordering into
dependency levels. Ghost cells can therefore be reconstructed sequentially
from lower to higher dependency levels without iterative ghost-cell updates.
This feature is particularly relevant to distributed-memory implementations,
where repeated reconstruction sweeps may also require repeated
inter-processor communication.

Alternative approaches have sought to retain narrow reconstruction stencils
under challenging grid and interface configurations. Picot and
Glockner~\cite{Picot:2018}, for example, introduced a virtual-node
reconstruction for anisotropic grids that retains a narrow stencil but
requires an additional geometrical construction. However, the numerical stability of the resulting reconstruction was not examined. These considerations motivate treating direct reconstruction, stencil compactness, ghost-cell dependencies, and reconstruction stability within a common framework.

Hybrid ghost cells were introduced in \cite{RapakaS:2018} in the context of
linear scalar wave propagation to satisfy reconstruction-stability constraints
while retaining a narrow Cartesian stencil. Classical ghost cells (CGC) have
centers strictly inside the solid region, whereas HGC centers may lie on either
side of the embedded boundary and therefore remain closer to the interface.
Here, we examine whether the stability criterion established from the earlier
linear analysis remains a useful indicator of reconstruction stability for
nonlinear incompressible Navier--Stokes simulations. CGC, HGC, and an
extended-stencil classical formulation denoted CGC\_ES are compared within the
same solver framework to separate the effects of ghost-cell placement,
reconstruction stability, and stencil compactness. Comparison between CGC and
CGC\_ES isolates the effect of satisfying the stability criterion, whereas
comparison between CGC\_ES and HGC highlights the benefit of achieving the same stability with a compact nearest-neighbour stencil.

The resulting framework therefore combines four features: direct analytical
boundary reconstruction, non-iterative dependency ordering,
stability-preserving ghost-cell placement with a compact stencil, and direct
evaluation of spatial gradients on the embedded boundary from the same
reconstruction polynomial. The latter enables pressure gradients, wall shear
stresses, drag, and lift to be evaluated without a separate
surface-reconstruction or filtering procedure.

The remainder of the article is organized as follows.
Section~\ref{num_method} summarizes the governing equations and numerical
method. Section~\ref{sec:sort} describes the identification of classical and
hybrid ghost cells and introduces the dependency ordering used for
non-iterative reconstruction. Sections~\ref{sec:bilinear} and
\ref{sec:trilinear} present the analytical reconstruction expressions in two
and three dimensions, respectively. Section~\ref{sec:surface_quantities}
describes the analytical evaluation of solution values and spatial derivatives
directly on the embedded boundary and their use in computing pressure forces,
wall stresses, drag, and lift. Section~\ref{sec:extended_reconstruction}
introduces the extended-stencil classical ghost-cell formulation CGC\_ES.
The stability behavior of CGC, CGC\_ES, and HGC is examined in
Sec.~\ref{sec:stability_results} using flow past a circular cylinder and an
airfoil. Finally, Sec.~\ref{sec:conclusions} summarizes the main conclusions.

\section{Governing equations and numerical method}\label{num_method}
Three dimensional Navier-Stokes equations for incompressible flow with the Boussinesq approximation for density coupling are solved numerically on a Cartesian grid along with an immersed boundary method to handle complex geometries in the flow field \cite{RapakaS:2016}. Second order central discretization is used for spatial derivatives on a colocated grid. A mixed RK3-ADI marching scheme is used to advance the solution in time \cite{RapakaS:2016} wherein the diffusion terms are advanced implicitly using an ADI method ~ \citep{Douglas_62,pozrikidis_book:2008} and all other terms are advanced using an explicit, low-storage Runge-Kutta-Wray3 (RKW3) time marching scheme. The solver allows the use of inhomogeneous boundary conditions in all three directions.
Further details on the solution algorithm are available in \cite{RapakaS:2016}.
\begin{eqnarray}
 \delux{\ui}{x_i} &=& 0\\
 \delux{\ui}{t} + \delux{(U_j u_i)}{x_j} &=& -\frac{1}{\rho_0}\delux{p}{x_i} + \nu \delsux{\ui}{x_j} - g \frac{\rs}{\rho_0} \delta_{i3} - \delux{\tau_{ij}}{x_j} \; , \\
 \delux{\rs}{t} + \delux{(U_j \rs)}{x_j} &=&  \kappa \delsux{\rs}{x_j} - u_i \delux{\rb}{x_i} \delta_{i3} - \delux{\lambda_{j}}{x_j} \; , 
\end{eqnarray}
where $\ui$ and $p$ are the velocity and pressure colocated at the cell center. $\rs (x,y,z,t)$ is the density perturbation relative to the background density $\rb (z)$ which is computed at the cell center $P$. $U_i$ are the face velocity components defined at the cell faces which are offset relative to the cell center by half the cell width in each direction. $U_i$ are used in the advection and Poisson source terms to avoid pressure-velocity decoupling (see \cite{RapakaS:2016} for details). $\tau_{ij}$ and $\lambda_{j}$ are the sub-grid scale stresses for momentum and density equations used in LES. The solver provides the option to estimate the sub-grid scale stresses using the standard Smagorinsky Model (SM) of \cite{Deardorff70} or the Dynamic Smagorinsky Model (DSM) of \cite{germano91}. However, the present work is focused on homogeneous, laminar flow simulations and the density equation, gravity term in the momentum equations, and the sub-grid scale models are not used here.

The solver employs a discrete immersed boundary method \cite{RapakaS:2016} wherein the governing equations are solved only inside the fluid domain and the solid region is ignored. A trilinear reconstruction is employed for ghost cells near the immersed boundary (see \cite{RapakaS:2016} or Sec. \ref{sec:trilinear} in this article). Notably, numerical computation of the reconstruction weights in \cite{RapakaS:2016} is replaced with the analytical expressions presented in this work (Sec. \ref{sec:trilinear}). Further, a non-iterative algorithm for ghost cell reconstruction, presented in Sec. \ref{sec:sort}, and an extended reconstruction stencil, presented in Sec. \ref{sec:extended_reconstruction}, that satisfies the stability criteria for anisotropic grids are implemented in the solver.
\section{Ghost-cell identification and dependency ordering}\label{sec:sort}
\subsection{Identification of ghost cells}\label{sec:def_HGC}
\begin{figure}[t]
 \centering
\begin{minipage}{0.45\linewidth}\centering
 \includegraphics[width=2.75in,trim={5.4cm 2.7cm 0.in 0.in},clip]{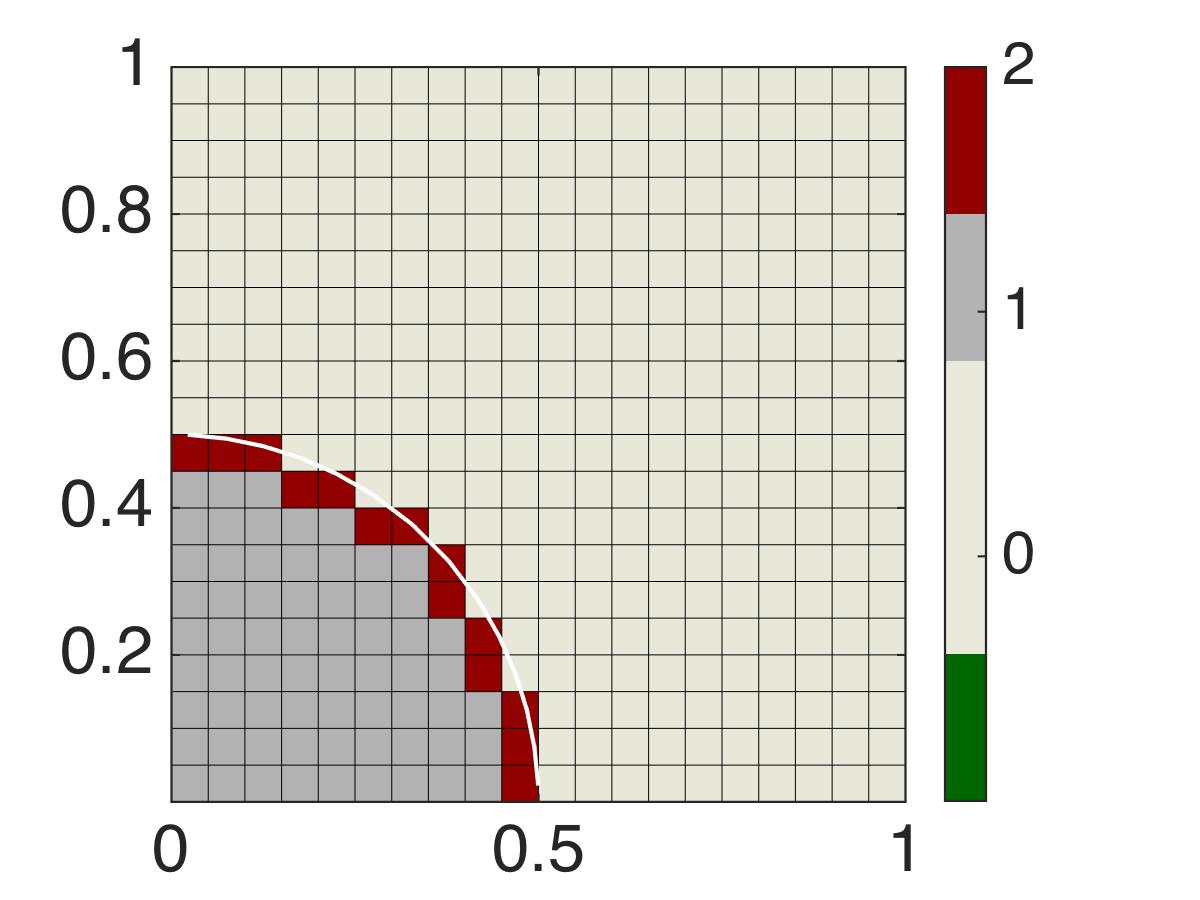} 
 \vspace*{-20ex}
 \begin{center}
     FLUID\\
     \vspace{10ex}
     \hspace{-10ex}
      SOLID
      \hspace{20ex}
      \vspace{-10ex}
\end{center}
 \vspace*{11ex}
 (a)
 \end{minipage}
 \centering
 \begin{minipage}{0.45\linewidth}\centering
  \includegraphics[width=2.75in,trim={5.4cm 2.7cm 0.in 0.in},clip]{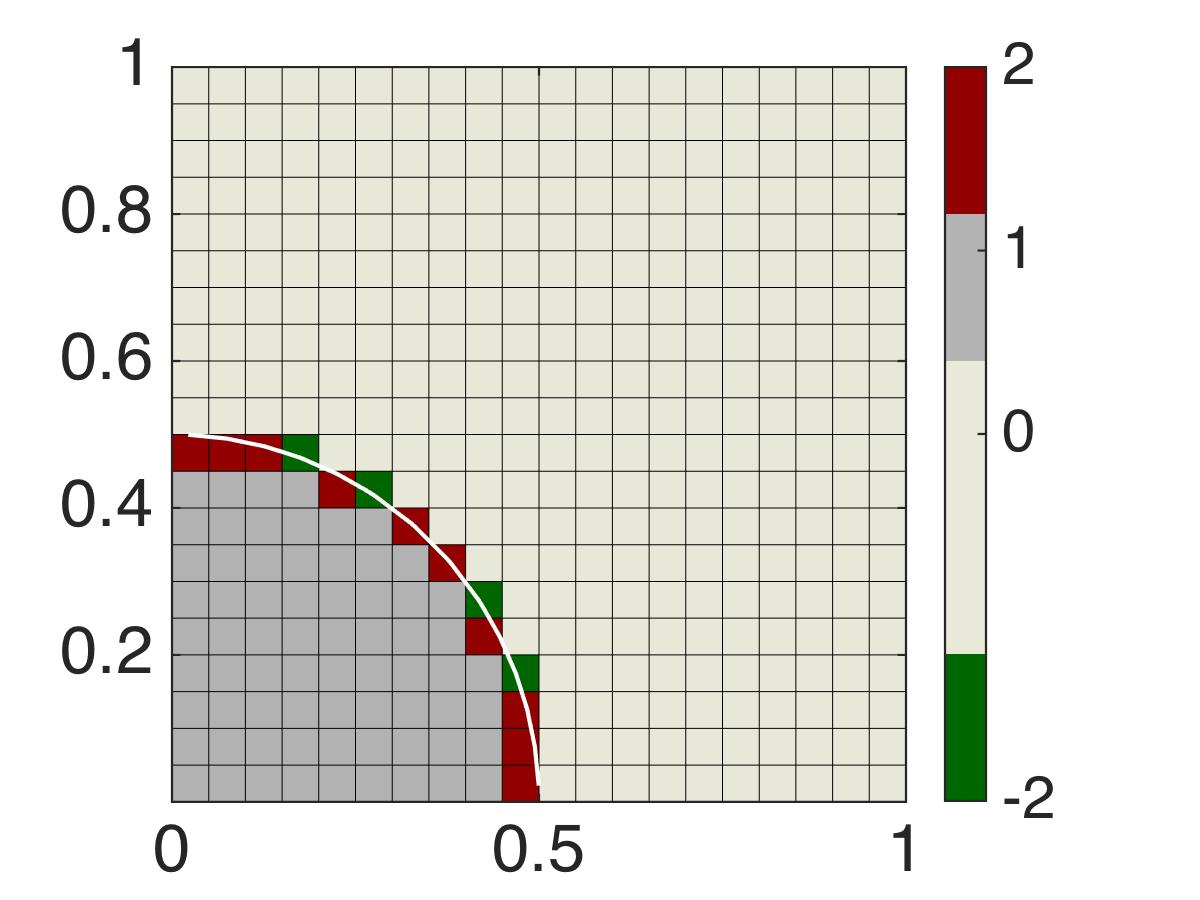} 
 \vspace*{-20ex}
 \begin{center}
     FLUID\\
     \vspace{10ex}
     \hspace{-10ex}
      SOLID
      \hspace{20ex}
      \vspace{-10ex}
\end{center}
 \vspace*{11ex}
(b)
 \end{minipage}
 \caption{Classification of ghost-cell strategies based on location relative to the embedded boundary, with fluid, solid, and ghost cells tagged as 0, 1, and $\pm 2$, respectively. (a) Classical ghost cells (CGC), which are strictly restricted to having their centers within the solid region (tagged $+2$) \citep{MittalDBNVL:2008}. (b) Hybrid ghost cells (HGC), which allow ghost-cell centers to reside in either the solid (tagged $+2$) or fluid (tagged $-2$) region depending on proximity to the interface (see Sec.~\ref{sec:def_HGC}).}
 \label{fig:gc_arrangement}
\end{figure}

First, all cells are classified as either fluid or solid according to the
location of their cell centers. Fluid and solid cells are assigned tags 0 and
1, respectively. In the classical ghost-cell (CGC) formulation, solid cells
immediately adjacent to the embedded boundary (EB) are designated as ghost
cells \citep{MittalDBNVL:2008}; see Fig.~\ref{fig:gc_arrangement}(a). In
contrast, the hybrid ghost-cell (HGC) classification introduced in
\cite{RapakaS:2018} allows ghost-cell centers to lie on either side of the EB,
as illustrated in Fig.~\ref{fig:gc_arrangement}(b). This placement facilitates
satisfaction of the reconstruction-stability criterion while retaining a
nearest-neighbour Cartesian stencil.

To identify HGC, we first define the characteristic cell thickness in the
boundary-normal direction as
\begin{equation}
\Delta n
=
\left[
\left(\frac{n_x}{\Delta x}\right)^2
+
\left(\frac{n_y}{\Delta y}\right)^2
+
\left(\frac{n_z}{\Delta z}\right)^2
\right]^{-1/2},
\label{eqn:delta_n}
\end{equation}
where
\begin{equation}
\hat{\boldsymbol{n}}
=
n_x\hat{\boldsymbol{i}}
+
n_y\hat{\boldsymbol{j}}
+
n_z\hat{\boldsymbol{k}}
\end{equation}
is the unit normal to the EB pointing into the fluid, and
$\Delta x$, $\Delta y$, and $\Delta z$ are the ghost cell ($G$) widths.
For a cell center $\boldsymbol{x}_c$, let $\boldsymbol{x}_s$ denote the
intersection of the EB with the boundary normal passing through
$\boldsymbol{x}_c$. We define the signed normal distance
\begin{equation}
\psi =
\begin{cases}
 \,\,\,\|\boldsymbol{x}_c-\boldsymbol{x}_s\|, & \boldsymbol{x}_c
 \text{ in the solid},\\[2mm]
-\|\boldsymbol{x}_c-\boldsymbol{x}_s\|, & \boldsymbol{x}_c
 \text{ in the fluid}.
\end{cases}
\label{eqn:signed_distance}
\end{equation}
Thus, $\psi\geq0$ in the solid and $\psi<0$ in the fluid. A cell is initially
classified as an HGC if
\begin{equation}
\begin{cases}
\,\,\,\psi \leq \Delta n/2, & \boldsymbol{x}_c \text{ in the solid},\\[1mm]
-\psi < \Delta n/2, & \boldsymbol{x}_c \text{ in the fluid}.
\end{cases}
\label{eqn:HGC_criterion}
\end{equation}

Following the initial tagging, some candidate ghost cells whose centers lie
in the fluid region may have no solid-cell neighbour along any Cartesian
direction. Likewise, some candidate ghost cells whose centers lie in the
solid region may have no fluid-cell neighbour along any Cartesian direction.
These excess candidates are reclassified as fluid and solid cells,
respectively. This reclassification maximizes the number of cells retained as fluid cells
and removes solid-side ghost cells that do not appear in the computational
stencil of the fluid cells. The HGC configuration defined above
inherently avoids ill-conditioned cases in the bilinear and trilinear
reconstruction procedures described in Secs.~\ref{sec:bilinear} and
\ref{sec:trilinear}.
\begin{figure}[t]
\centering

\begin{minipage}{0.48\linewidth}
\centering
\begin{tikzpicture}[scale=0.82, every node/.style={font=\scriptsize}]

\def\nodesize{5.8pt}
\def\ghostsize{8pt}

\fill[gray!16]
(0,0) -- (9,0) -- (9,6.25)
.. controls (8.2,6.05) and (7.2,5.85) .. (6.3,5.65)
.. controls (5.4,5.35) and (4.8,4.75) .. (4.2,4.25)
.. controls (3.2,3.65) and (2.0,3.45) .. (0,3.25)
-- cycle;

\draw[step=1.0, gray!55, very thin] (0,0) grid (9,7);

\draw[black, thick]
(0,3.25)
.. controls (2.0,3.45) and (3.2,3.65) .. (4.2,4.25)
.. controls (4.8,4.75) and (5.4,5.35) .. (6.3,5.65)
.. controls (7.2,5.85) and (8.2,6.05) .. (9,6.25);

\node at (1.05,0.85) {SOLID};
\node at (1.05,6.15) {FLUID};

\foreach \x/\y in {
0.5/3.5,
0.5/4.5,1.5/4.5,2.5/4.5,3.5/4.5,
0.5/5.5,1.5/5.5,2.5/5.5,3.5/5.5,4.5/5.5,5.5/5.5,
0.5/6.5,1.5/6.5,2.5/6.5,3.5/6.5,4.5/6.5,5.5/6.5,6.5/6.5,7.5/6.5,8.5/6.5}
{
  \node[circle, draw=black, fill=white, minimum size=\nodesize, inner sep=0pt] at (\x,\y) {};
}

\foreach \x/\y in {
0.5/0.5,1.5/0.5,2.5/0.5,3.5/0.5,4.5/0.5,5.5/0.5,6.5/0.5,7.5/0.5,8.5/0.5,
0.5/1.5,1.5/1.5,2.5/1.5,3.5/1.5,4.5/1.5,5.5/1.5,6.5/1.5,7.5/1.5,8.5/1.5,
1.5/2.5,2.5/2.5,3.5/2.5,4.5/2.5,5.5/2.5,6.5/2.5,7.5/2.5,8.5/2.5,
4.5/3.5,5.5/3.5,6.5/3.5,7.5/3.5,8.5/3.5,
6.5/4.5,7.5/4.5,8.5/4.5}
{
  \node[rectangle, draw=black, fill=white, minimum size=\nodesize, inner sep=0pt] at (\x,\y) {};
}


\node[rectangle, draw=black, fill=blue!25, minimum size=\ghostsize, inner sep=0pt] (L1d) at (0.5,2.5) {};
\node[below right=-1pt] at (L1d) {$L_1$};

\node[rectangle, draw=black, fill=blue!25, minimum size=\ghostsize, inner sep=0pt] (L1a) at (1.5,3.5) {};
\node[below right=-1pt] at (L1a) {$L_1$};

\node[rectangle, draw=black, fill=green!30, minimum size=\ghostsize, inner sep=0pt] (L2a) at (2.5,3.5) {};
\node[below right=-1pt] at (L2a) {$L_2$};

\node[rectangle, draw=black, fill=orange!40, minimum size=\ghostsize, inner sep=0pt] (L3a) at (3.5,3.5) {};
\node[below right=-1pt] at (L3a) {$L_3$};

\node[rectangle, draw=black, fill=blue!25, minimum size=\ghostsize, inner sep=0pt] (L1c) at (4.5,4.5) {};
\node[below right=-1pt] at (L1c) {$L_1$};

\node[rectangle, draw=black, fill=green!30, minimum size=\ghostsize, inner sep=0pt] (L2b) at (5.5,4.5) {};
\node[below right=-1pt] at (L2b) {$L_2$};

\node[rectangle, draw=black, fill=blue!25, minimum size=\ghostsize, inner sep=0pt] (L2c) at (6.5,5.5) {};
\node[below right=-1pt] at (L2c) {$L_1$};

\node[rectangle, draw=black, fill=green!30, minimum size=\ghostsize, inner sep=0pt] (L3b) at (7.5,5.5) {};
\node[below right=-1pt] at (L3b) {$L_2$};

\node[rectangle, draw=black, fill=orange!40, minimum size=\ghostsize, inner sep=0pt] (L4b) at (8.5,5.5) {};
\node[below right=-1pt] at (L4b) {$L_3$};


\draw[blue!70, thick, dashed, rounded corners]
  (0.08,3.08) rectangle (1.92,4.92);
\node[blue!70, anchor=west] at (0.98,5.20) {$\chi(L_1)$};

\draw[blue!70, thick, dashed, rounded corners]
  (3.08,4.08) rectangle (4.92,5.92);

\draw[blue!70, thick, dashed, rounded corners]
  (5.08,5.08) rectangle (6.92,6.92);

\draw[green!60!black, thick, dashed, rounded corners]
  (1.08,3.08) rectangle (2.92,4.92);
\node[green!60!black, anchor=west] at (1.98,5.20) {$\chi(L_2)$};

\draw[green!60!black, thick, dashed, rounded corners]
  (4.08,4.08) rectangle (5.92,5.92);

\draw[green!60!black, thick, dashed, rounded corners]
  (6.08,5.08) rectangle (7.92,6.92);

\draw[orange!80!black, thick, dashed, rounded corners]
  (2.08,3.08) rectangle (3.92,4.92);
\node[orange!80!black, anchor=west] at (2.98,5.20) {$\chi(L_3)$};

\draw[orange!80!black, thick, dashed, rounded corners]
  (7.08,5.08) rectangle (8.92,6.92);


\draw[->, thick, blue!70] (0.5,3.5) -- (L1a);
\draw[->, thick, blue!70] (0.5,4.5) -- (L1a);
\draw[->, thick, blue!70] (1.5,4.5) -- (L1a);

\draw[->, thick, green!60!black] (L1a) -- (L2a);
\draw[->, thick, green!60!black] (L2c) -- (L3b);
\draw[->, thick, green!60!black] (L1c) -- (L2b);

\draw[->, thick, orange!80!black] (L2a) -- (L3a);
\draw[->, thick, orange!80!black] (L3b) -- (L4b);


\draw[->, thick, green!60] (1.5,4.5) -- (L2a);
\draw[->, thick, green!60] (1.5,3.5) -- (L2a);
\draw[->, thick, green!60] (2.5,4.5) -- (L2a);

\draw[->, thick, orange!80] (2.5,4.5) -- (L3a);
\draw[->, thick, orange!80] (2.5,3.5) -- (L3a);
\draw[->, thick, orange!80] (3.5,4.5) -- (L3a);

\node[circle, draw=black, fill=white, minimum size=\nodesize, inner sep=0pt] at (0.45,-0.45) {};
\node[anchor=west] at (0.65,-0.45) {fluid};

\node[rectangle, draw=black, fill=white, minimum size=\nodesize, inner sep=0pt] at (1.55,-0.45) {};
\node[anchor=west] at (1.75,-0.45) {solid};

\node[rectangle, draw=black, fill=blue!25, minimum size=7pt, inner sep=0pt] at (2.70,-0.45) {};
\node[anchor=west] at (2.90,-0.45) {$L_1$};

\node[rectangle, draw=black, fill=green!30, minimum size=7pt, inner sep=0pt] at (3.55,-0.45) {};
\node[anchor=west] at (3.75,-0.45) {$L_2$};

\node[rectangle, draw=black, fill=orange!40, minimum size=7pt, inner sep=0pt] at (4.40,-0.45) {};
\node[anchor=west] at (4.60,-0.45) {$L_3$};


\end{tikzpicture}

\vspace{0.3em}
{\scriptsize (a) Ghost cells grouped into dependency levels}
\end{minipage}
\hfill
\begin{minipage}{0.48\linewidth}
\centering
\begin{tikzpicture}[
    node distance=1.2cm and 1.4cm,
    every node/.style={font=\scriptsize},
    fluid/.style={circle, draw=black, fill=white, minimum size=6mm},
    ghost/.style={circle, draw=black, fill=blue!15, minimum size=6mm},
    edge/.style={->, thick}
]

\node[anchor=east] at (-1.2,-1.4) {Level $L_1$};
\node[anchor=east] at (-1.2,-2.8) {Level $L_2$};
\node[anchor=east] at (-1.2,-4.2) {Level $L_3$};
\node[anchor=east] at (-1.2,-5.6) {Level $L_n$};


\node[ghost] (a1) at (0.6,-1.4) {$G_1$};
\node[ghost] (a2) at (2.4,-1.4) {$G_2$};
\node[ghost] (a3) at (4.2,-1.4) {$G_3$};

\node[ghost] (b1) at (1.2,-2.8) {$G_4$};
\node[ghost] (b2) at (3.6,-2.8) {$G_5$};

\node[ghost] (c1) at (2.4,-4.2) {$G_6$};
\node[ghost] (c2) at (4.2,-4.2) {$G_7$};

\node[ghost] (d1) at (3.3,-5.6) {$G_m$};


\draw[edge] (a1) -- (b1);
\draw[edge] (a2) -- (b1);
\draw[edge] (a2) -- (b2);
\draw[edge] (a3) -- (b2);

\draw[edge] (b1) -- (c1);
\draw[edge] (b2) -- (c1);
\draw[edge] (b2) -- (c2);

\draw[edge] (c1) -- (d1);
\draw[edge] (c2) -- (d1);

\draw[dashed, gray] (-0.4,-0.7) -- (5.2,-0.7);
\draw[dashed, gray] (-0.4,-2.1) -- (5.2,-2.1);
\draw[dashed, gray] (-0.4,-3.5) -- (5.2,-3.5);
\draw[dashed, gray] (-0.4,-4.9) -- (5.2,-4.9);

\node[align=center, font=\scriptsize] at (2.25,-0.3)
{Topological ordering of ghost cells \\
$L_1 \rightarrow L_2 \rightarrow L_3 \rightarrow \cdots \rightarrow L_n$};


\end{tikzpicture}

\vspace{0.3em}
{\scriptsize (b) Directed acyclic dependency graph}
\end{minipage}

\caption{
Dependency ordering of ghost cells.
(a) Representative embedded-boundary configuration showing ghost cells classified into dependency levels according to reconstruction dependencies.
(b) Directed acyclic graph associated with the reconstruction procedure.
Edges indicate reconstruction dependencies. Ghost cells belonging to level $L_k$ depend only on fluid nodes and ghost cells in lower levels $L_1,\ldots,L_{k-1}$.
The graph admits a topological ordering, permitting direct level-by-level reconstruction without iterative updates.
}
\label{fig:dependency_ordering}
\end{figure}

\narsimha{
\subsection{Topological ordering of ghost-cell dependencies}
This section introduces a non-iterative ghost-cell reconstruction procedure
based on the dependency structure of the reconstruction stencil. In
conventional embedded-boundary methods, the reconstruction stencil of a ghost
cell may contain neighboring ghost cells whose values are themselves unknown,
and this coupling is typically resolved through iterative reconstruction
procedures. For the reconstruction stencil considered here, however, the
ghost-cell dependencies generated by the Cartesian neighbour selection form a
directed acyclic graph. Accordingly, the ghost cells can be arranged in a
topological ordering such that every ghost cell depends only on fluid nodes or
on ghost cells appearing earlier in the ordering. This structure is
illustrated schematically in Fig.~\ref{fig:dependency_ordering}.

The topological ordering is implemented by partitioning the ghost cells into
dependency levels $L_1,L_2,\ldots,L_n$. Level $L_1$ contains ghost cells whose
reconstruction stencils contain only fluid nodes. After these values are
available, level $L_2$ contains cells whose stencils involve only fluid nodes
and ghost cells in $L_1$. More generally, a ghost cell is assigned to level
$L_k$ once all ghost-cell values required by its reconstruction belong to
levels $L_1,\ldots,L_{k-1}$. Reconstruction can therefore proceed directly
from $L_1$ through $L_n$ without iterative ghost-cell updates, as illustrated
in Fig.~\ref{fig:dependency_ordering}. Occasionally, one or more components of
the embedded-boundary normal vector satisfy $n_x=0$, $n_y=0$, or $n_z=0$. In
such cases, the corresponding reconstruction weights vanish (see
Secs.~\ref{sec:bilinear} and \ref{sec:trilinear}), eliminating dependence on
neighboring ghost cells in those directions. The affected ghost cells are
therefore assigned to the appropriate dependency level $L_k$.

For implementation, ghost-cell indices are stored according to their
dependency level, from $L_1$ through $L_n$. The ordering of ghost cells within
a given level is arbitrary because cells belonging to the same level are
mutually independent. The maximum number of dependency levels also provides a
measure of the propagation depth encountered by an iterative reconstruction
procedure. For a dependency chain of depth $n$, information from the known
fluid nodes must propagate successively through $L_1,L_2,\ldots,L_n$. A
conventional synchronous reconstruction therefore requires at least $n$
sweeps for updated information to propagate through a dependency chain of
depth $n$. Additional iterations may be required in practice to achieve
convergence; for example, Mittal et al.~\cite{MittalDBNVL:2008} employed a
coupled Gauss--Seidel procedure requiring approximately ten iterations.

Once the ordering has been constructed, ghost cells are reconstructed
sequentially from $L_1$ to $L_n$, with each ghost cell evaluated only after
the quantities required by its stencil are available. Thus, each ghost cell
is reconstructed once rather than being repeatedly updated through iterative
sweeps. The ordering procedure scales linearly with the number of ghost cells,
$\mathcal{O}(N_g)$.

In distributed-memory implementations, ghost-cell dependencies may span MPI
subdomain boundaries and therefore require inter-processor communication. The
topological ordering does not eliminate this communication, but localizes it
to the dependency level that has just been reconstructed. After completing
level $L_k$, only the updated ghost cells in $L_k$ that participate in
reconstruction stencils on neighboring subdomains need to be exchanged before
proceeding to level $L_{k+1}$. In contrast, a conventional iterative
reconstruction updates the full set of ghost cells during each sweep and
typically performs the usual halo exchange of processor-boundary data after
each iteration. This can involve communication of substantially more data than
the subset of level-$L_k$ ghost-cell values actually required by neighboring
subdomains. The dependency ordering therefore replaces repeated global
reconstruction sweeps with level-by-level reconstruction and communication.


\subsection{Comparison of the ghost cell reconstruction strategies}
\begin{figure}[htp]
\centering

\begin{minipage}{0.33\textwidth}
\centering
\begin{tikzpicture}[scale=0.85,>=stealth]

\def\R{3.75}

\fill[gray!18]
(0,0) --
(\R,0)
arc[start angle=0,end angle=90,radius=\R]
-- cycle;

\draw[step=1.0,dashed,gray!65] (0,0) grid (6,6);

\draw[very thick]
(\R,0)
arc[start angle=0,end angle=90,radius=\R];

\node at (5.15,5.05) {\scriptsize FLUID};
\node at (1.05,0.85) {\scriptsize SOLID};

\draw[red!75!black,dotted,thick]
(3.25,0) arc[start angle=0,end angle=90,radius=3.25];

\draw[red!75!black,dotted,thick]
(4.25,0) arc[start angle=0,end angle=90,radius=4.25];


\fill[blue!18,opacity=0.65]
(0.5,3.5) rectangle (1.5,4.5);
\draw[blue!70!black,thick]
(0.5,3.5) rectangle (1.5,4.5);

\fill[blue!18,opacity=0.65]
(1.5,3.5) rectangle (2.5,4.5);
\draw[blue!70!black,thick]
(1.5,3.5) rectangle (2.5,4.5);

\fill[blue!18,opacity=0.65]
(2.5,2.5) rectangle (3.5,3.5);
\draw[blue!70!black,thick]
(2.5,2.5) rectangle (3.5,3.5);

\fill[blue!18,opacity=0.65]
(3.5,1.5) rectangle (4.5,2.5);
\draw[blue!70!black,thick]
(3.5,1.5) rectangle (4.5,2.5);

\fill[blue!18,opacity=0.65]
(3.5,0.5) rectangle (4.5,1.5);
\draw[blue!70!black,thick]
(3.5,0.5) rectangle (4.5,1.5);


\foreach \x/\y in {
0.5/3.5,0.5/4.5,0.5/5.5,
1.5/3.5,1.5/4.5,1.5/5.5,
2.5/3.5,2.5/4.5,2.5/5.5,
3.5/1.5,3.5/2.5,3.5/3.5,3.5/4.5,3.5/5.5,
4.5/0.5,4.5/1.5,4.5/2.5,4.5/3.5,4.5/4.5,4.5/5.5,
5.5/0.5,5.5/1.5,5.5/2.5,5.5/3.5,5.5/4.5,5.5/5.5
}{
\draw[thick] (\x,\y) circle (0.10);
}


\foreach \x/\y in {
0.5/0.5,0.5/1.5,0.5/2.5,
1.5/0.5,1.5/1.5,
2.5/0.5
}{
\draw[thick]
(\x-0.10,\y-0.10)
rectangle
(\x+0.10,\y+0.10);
}


\foreach \x/\y in {
0.5/3.5,
1.5/2.5,
2.5/1.5,
2.5/2.5,
3.5/0.5
}{
\fill
(\x-0.11,\y-0.11)
rectangle
(\x+0.11,\y+0.11);
}


\coordinate (G1)  at (0.5,3.5);
\coordinate (S1)  at (0.5310,3.7122);
\coordinate (IP1) at (0.5620,3.9244);

\coordinate (G2)  at (1.5,2.5);
\coordinate (S2)  at (1.9290,3.2150);
\coordinate (IP2) at (2.3580,3.9300);

\coordinate (G3)  at (2.5,2.5);
\coordinate (S3)  at (2.660,2.630);
\coordinate (IP3) at (2.820,2.760);

\coordinate (G4)  at (2.5,1.5);
\coordinate (S4)  at (3.2150,1.9290);
\coordinate (IP4) at (3.9300,2.3580);

\coordinate (G5)  at (3.5,0.5);
\coordinate (S5)  at (3.7,0.550);
\coordinate (IP5) at (3.9,0.60);


\draw[thick] (G1)--(S1)--(IP1);
\draw[thick] (G2)--(S2)--(IP2);
\draw[thick] (G3)--(S3)--(IP3);
\draw[thick] (G4)--(S4)--(IP4);
\draw[thick] (G5)--(S5)--(IP5);


\foreach \p in {S1,S2,S3,S4,S5}{
\fill[red!75!black] (\p) circle (0.11);
\draw[thick,red!75!black] (\p) circle (0.11);
}


\foreach \p in {IP1,IP2,IP3,IP4,IP5}{
\fill[green!65!black] (\p) circle (0.095);
\draw[thick,green!45!black] (\p) circle (0.095);
}




\node[below left] at (G1)
{\scriptsize $G_1$};
\node[above left,inner sep=1pt] at (S1)
{\scriptsize $S_1$};
\node[above right,inner sep=1pt] at (IP1)
{\scriptsize $IP_1$};

\node[below left] at (G2)
{\scriptsize $G_2$};
\node[above left,inner sep=1pt] at (S2)
{\scriptsize $S_2$};
\node[above right,inner sep=1pt] at (IP2)
{\scriptsize $IP_2$};

\node[below left] at (G3)
{\scriptsize $G_3$};
\node[above left,inner sep=1pt] at (S3)
{\scriptsize $S_3$};
\node[above right,inner sep=1pt] at (IP3)
{\scriptsize $IP_3$};

\node[below left] at (G4)
{\scriptsize $G_4$};
\node[above left,inner sep=1pt] at (S4)
{\scriptsize $S_4$};
\node[above right,inner sep=1pt] at (IP4)
{\scriptsize $IP_4$};

\node[below left] at (G5)
{\scriptsize $G_5$};
\node[above left,inner sep=1pt] at (S5)
{\scriptsize $S_5$};
\node[above right,inner sep=1pt] at (IP5)
{\scriptsize $IP_5$};


\begin{scope}[shift={(0.20,-0.55)}]

\draw[thick] (0,0) circle (0.10);
\node[right] at (0.18,0)
{\scriptsize fluid};

\draw[thick]
(1.25,-0.10)
rectangle
(1.45,0.10);
\node[right] at (1.60,0)
{\scriptsize solid};

\fill
(2.50,-0.10)
rectangle
(2.70,0.10);
\node[right] at (2.85,0)
{\scriptsize ghost};

\fill[red!75!black]
(4.05,0)
circle (0.08);
\node[right] at (4.20,0)
{\scriptsize $S$};

\fill[green!65!black]
(5.10,0)
circle (0.08);
\node[right] at (5.25,0)
{\scriptsize IP};

\fill[blue!18,opacity=0.65]
(-0.1,-0.30)
rectangle
(0.2,-0.54);

\draw[blue!70!black,thick]
(-0.1,-0.30)
rectangle
(0.2,-0.54);

\node[right] at (0.22,-0.4)
{\scriptsize IP stencil};

\end{scope}

\end{tikzpicture}
{\scriptsize \\(a) Image-point reconstruction of CGC ~\cite{MittalDBNVL:2008}}
\end{minipage}
\begin{minipage}{0.33\textwidth}
\centering
\begin{tikzpicture}[scale=0.85,>=stealth]

\def\R{3.75}

\fill[gray!18]
(0,0) --
(\R,0)
arc[start angle=0,end angle=90,radius=\R]
-- cycle;

\draw[step=1.0,dashed,gray!65] (0,0) grid (6,6);

\draw[very thick]
(\R,0)
arc[start angle=0,end angle=90,radius=\R];

\node at (5.15,5.05) {\scriptsize FLUID};
\node at (1.05,0.85) {\scriptsize SOLID};

\draw[red!75!black,dotted,thick]
(3.25,0) arc[start angle=0,end angle=90,radius=3.25];

\draw[red!75!black,dotted,thick]
(4.25,0) arc[start angle=0,end angle=90,radius=4.25];


\fill[blue!18,opacity=0.65]
(0.5,3.5) rectangle (1.5,4.5);
\draw[blue!70!black,thick]
(0.5,3.5) rectangle (1.5,4.5);

\fill[blue!18,opacity=0.65]
(1.5,2.5) rectangle (2.5,3.5);
\draw[blue!70!black,thick]
(1.5,2.5) rectangle (2.5,3.5);

\fill[blue!18,opacity=0.65]
(2.5,2.5) rectangle (3.5,3.5);
\draw[blue!70!black,thick]
(2.5,2.5) rectangle (3.5,3.5);

\fill[blue!18,opacity=0.65]
(2.5,1.5) rectangle (3.5,2.5);
\draw[blue!70!black,thick]
(2.5,1.5) rectangle (3.5,2.5);

\fill[blue!18,opacity=0.65]
(3.5,0.5) rectangle (4.5,1.5);
\draw[blue!70!black,thick]
(3.5,0.5) rectangle (4.5,1.5);


\foreach \x/\y in {
0.5/3.5,0.5/4.5,0.5/5.5,
1.5/3.5,1.5/4.5,1.5/5.5,
2.5/3.5,2.5/4.5,2.5/5.5,
3.5/1.5,3.5/2.5,3.5/3.5,3.5/4.5,3.5/5.5,
4.5/0.5,4.5/1.5,4.5/2.5,4.5/3.5,4.5/4.5,4.5/5.5,
5.5/0.5,5.5/1.5,5.5/2.5,5.5/3.5,5.5/4.5,5.5/5.5
}{
\draw[thick] (\x,\y) circle (0.10);
}


\foreach \x/\y in {
0.5/0.5,0.5/1.5,0.5/2.5,
1.5/0.5,1.5/1.5,
2.5/0.5
}{
\draw[thick]
(\x-0.10,\y-0.10)
rectangle
(\x+0.10,\y+0.10);
}


\foreach \x/\y in {
0.5/3.5,
1.5/2.5,
2.5/1.5,
2.5/2.5,
3.5/0.5
}{
\fill
(\x-0.11,\y-0.11)
rectangle
(\x+0.11,\y+0.11);
}


\coordinate (G1) at (0.5,3.5);
\coordinate (S1) at (0.5310,3.7122);

\coordinate (G2) at (1.5,2.5);
\coordinate (S2) at (1.9290,3.2150);

\coordinate (G3)  at (2.5,2.5);
\coordinate (S3)  at (2.660,2.630);

\coordinate (G4) at (2.5,1.5);
\coordinate (S4) at (3.2150,1.9290);

\coordinate (G5)  at (3.5,0.5);
\coordinate (S5)  at (3.7,0.550);

\draw[thick] (G1)--(S1);
\draw[thick] (G2)--(S2);
\draw[thick] (G3)--(S3);
\draw[thick] (G4)--(S4);
\draw[thick] (G5)--(S5);

\foreach \p in {S1,S2,S3,S4,S5}{
\fill[red!75!black] (\p) circle (0.11);
\draw[thick,red!75!black] (\p) circle (0.11);
}


\node[below left] at (G1) {\scriptsize $G_1$};
\node[above left,inner sep=1pt] at (S1) {\scriptsize $S_1$};

\node[below left] at (G2) {\scriptsize $G_2$};
\node[above left,inner sep=1pt] at (S2) {\scriptsize $S_2$};

\node[below left] at (G3) {\scriptsize $G_3$};
\node[above left,inner sep=1pt] at (S3) {\scriptsize $S_3$};

\node[below left] at (G4) {\scriptsize $G_4$};
\node[above left,inner sep=1pt] at (S4) {\scriptsize $S_4$};

\node[below left] at (G5) {\scriptsize $G_5$};
\node[above left,inner sep=1pt] at (S5) {\scriptsize $S_5$};

\begin{scope}[shift={(0.20,-0.55)}]

\draw[thick] (0,0) circle (0.10);
\node[right] at (0.18,0) {\scriptsize fluid};

\draw[thick] (1.25,-0.10) rectangle (1.45,0.10);
\node[right] at (1.60,0) {\scriptsize solid};

\fill (2.50,-0.10) rectangle (2.70,0.10);
\node[right] at (2.85,0) {\scriptsize ghost};

\fill[red!75!black] (4.05,0) circle (0.08);
\node[right] at (4.20,0) {\scriptsize $S$};

\fill[blue!18,opacity=0.65] (-0.1,-0.30) rectangle (0.2,-0.54);
\draw[blue!70!black,thick] (-0.1,-0.30) rectangle (0.2,-0.54);
\node[right] at (0.22,-0.4) {\scriptsize direct stencil};

\end{scope}

\end{tikzpicture}
{\scriptsize \\(b) Direct reconstruction of CGC}
\end{minipage}
\begin{minipage}{0.33\textwidth}
\centering
\begin{tikzpicture}[scale=0.85,>=stealth]

\def\R{3.75}

\fill[gray!18]
(0,0) --
(\R,0)
arc[start angle=0,end angle=90,radius=\R]
-- cycle;

\draw[step=1.0,dashed,gray!65] (0,0) grid (6,6);

\draw[very thick]
(\R,0)
arc[start angle=0,end angle=90,radius=\R];

\node at (5.15,5.05) {\scriptsize FLUID};
\node at (1.05,0.85) {\scriptsize SOLID};

\draw[red!75!black,dotted,thick]
(3.25,0) arc[start angle=0,end angle=90,radius=3.25];

\draw[red!75!black,dotted,thick]
(4.25,0) arc[start angle=0,end angle=90,radius=4.25];



\fill[blue!18,opacity=0.65]
(0.5,3.5) rectangle (1.5,4.5);
\draw[blue!70!black,thick]
(0.5,3.5) rectangle (1.5,4.5);

\fill[blue!18,opacity=0.65]
(1.5,3.5) rectangle (2.5,4.5);
\draw[blue!70!black,thick]
(1.5,3.5) rectangle (2.5,4.5);

\fill[blue!18,opacity=0.65]
(2.5,2.5) rectangle (3.5,3.5);
\draw[blue!70!black,thick]
(2.5,2.5) rectangle (3.5,3.5);

\fill[blue!18,opacity=0.65]
(3.5,1.5) rectangle (4.5,2.5);
\draw[blue!70!black,thick]
(3.5,1.5) rectangle (4.5,2.5);

\fill[blue!18,opacity=0.65]
(3.5,0.5) rectangle (4.5,1.5);
\draw[blue!70!black,thick]
(3.5,0.5) rectangle (4.5,1.5);

\foreach \x/\y in {
0.5/4.5,0.5/5.5,
1.5/3.5,1.5/4.5,1.5/5.5,
2.5/3.5,2.5/4.5,2.5/5.5,
3.5/2.5,3.5/3.5,3.5/4.5,3.5/5.5,
4.5/0.5,4.5/1.5,4.5/2.5,4.5/3.5,4.5/4.5,4.5/5.5,
5.5/0.5,5.5/1.5,5.5/2.5,5.5/3.5,5.5/4.5,5.5/5.5
}{
\draw[thick] (\x,\y) circle (0.10);
}

\foreach \x/\y in {
0.5/0.5,0.5/1.5,0.5/2.5,
1.5/0.5,1.5/1.5,1.5/2.5,
2.5/0.5,2.5/1.5,2.5/2.5
}{
\draw[thick]
(\x-0.10,\y-0.10)
rectangle
(\x+0.10,\y+0.10);
}

\foreach \x/\y in {
0.5/3.5,
1.5/3.5,
2.5/2.5,
3.5/0.5,
3.5/1.5
}{
\fill
(\x-0.11,\y-0.11)
rectangle
(\x+0.11,\y+0.11);
}


\coordinate (G1) at (0.5,3.5);
\coordinate (S1) at (0.5310,3.7122);

\coordinate (G2) at (1.5,3.5);
\coordinate (S2) at (1.4779,3.4485);

\coordinate (G3)  at (2.5,2.5);
\coordinate (S3)  at (2.660,2.630);

\coordinate (G4) at (3.5,1.5);
\coordinate (S4) at (3.4485,1.4779);

\coordinate (G5)  at (3.5,0.5);
\coordinate (S5)  at (3.7,0.550);

\draw[thick] (G1)--(S1);
\draw[thick] (G2)--(S2);
\draw[thick] (G3)--(S3);
\draw[thick] (G4)--(S4);
\draw[thick] (G5)--(S5);

\foreach \p in {S1,S2,S3,S4,S5}{
\fill[red!75!black] (\p) circle (0.11);
\draw[thick,red!75!black] (\p) circle (0.11);
}


\node[below left] at (G1) {\scriptsize $G_1$};
\node[above left,inner sep=1pt] at (S1) {\scriptsize $S_1$};

\node[above right] at (G2) {\scriptsize $G_2$};
\node[below left,inner sep=1pt] at (S2) {\scriptsize $S_2$};

\node[below left] at (G3) {\scriptsize $G_3$};
\node[above left,inner sep=1pt] at (S3) {\scriptsize $S_3$};

\node[above right] at (G4) {\scriptsize $G_4$};
\node[below left,inner sep=1pt] at (S4) {\scriptsize $S_4$};

\node[below left] at (G5) {\scriptsize $G_5$};
\node[above left,inner sep=1pt] at (S5) {\scriptsize $S_5$};

\begin{scope}[shift={(0.20,-0.55)}]

\draw[thick] (0,0) circle (0.10);
\node[right] at (0.18,0) {\scriptsize fluid};

\draw[thick] (1.25,-0.10) rectangle (1.45,0.10);
\node[right] at (1.60,0) {\scriptsize solid};

\fill
(2.50,-0.10)
rectangle
(2.70,0.10);
\node[right] at (2.85,0) {\scriptsize HGC};

\fill[red!75!black] (4.05,0) circle (0.08);
\node[right] at (4.20,0) {\scriptsize $S$};

\fill[blue!18,opacity=0.65] (-0.1,-0.30) rectangle (0.2,-0.54);
\draw[blue!70!black,thick] (-0.1,-0.30) rectangle (0.2,-0.54);
\node[right] at (0.22,-0.4) {\scriptsize direct stencil};

\end{scope}

\end{tikzpicture}
{\scriptsize \\(c) Direct reconstruction of HGC}
\end{minipage}
\caption{
Comparison of representative ghost-cell reconstruction strategies for
embedded-boundary methods.
(a) Classical image-point reconstruction of Mittal et al.~\cite{MittalDBNVL:2008}, in which the solution is first reconstructed at an image point (IP) mirrored
across the embedded boundary and subsequently used to reconstruct the
ghost-cell value.
(b) Direct reconstruction using classical ghost cells (CGC). The ghost-cell
locations are the same as in the classical formulation, but the image-point
step is eliminated and the boundary condition is imposed directly through the
analytical reconstruction formula using the nearest-neighbour Cartesian stencil.
(c) Present hybrid ghost-cell (HGC) reconstruction, in which ghost-cell centers
may lie on either side of the embedded boundary and remain close to the
interface. This placement allows the stability criterion
Eq.~(\ref{eqn:stability_alpha_max}) to be satisfied while retaining the
nearest-neighbour Cartesian stencil.
}
\label{fig:comparison_reconstruction_methods}
\end{figure}
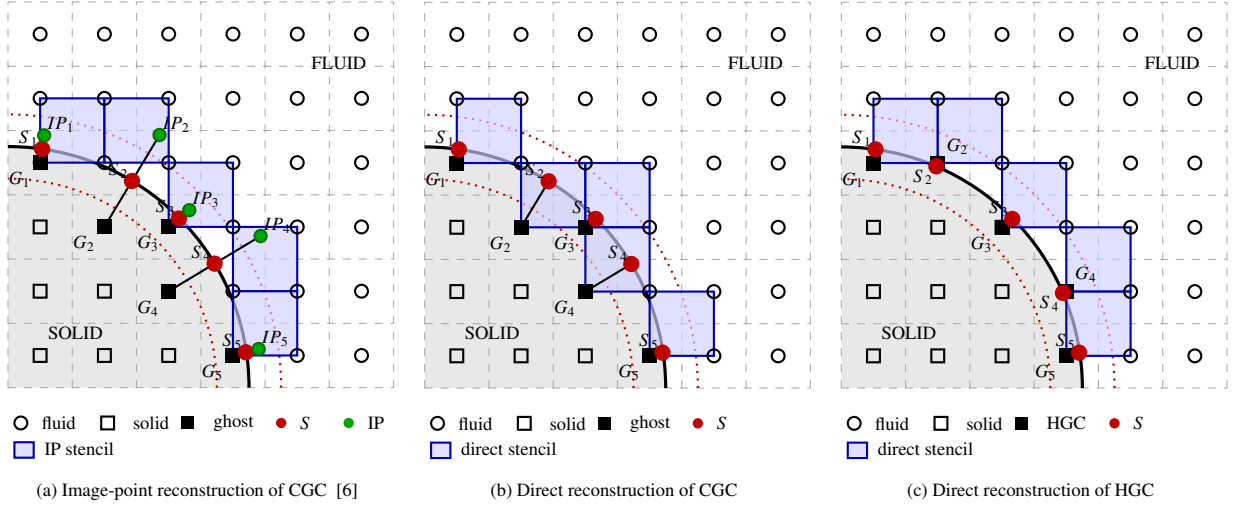
Figure~\ref{fig:comparison_reconstruction_methods} distinguishes the effects
of ghost-cell placement from those of the reconstruction procedure. In the
classical image-point method of Mittal et al.~\cite{MittalDBNVL:2008}, ghost
cells are restricted to the solid side of the embedded boundary and their
values are obtained indirectly: the solution is first reconstructed at a
mirrored image point in the fluid and is subsequently used to determine the
ghost-cell value, as illustrated in
Fig.~\ref{fig:comparison_reconstruction_methods}(a).

The CGC formulation used in the present work retains the same classical
ghost-cell placement but eliminates the image-point construction. Instead, the
boundary condition is imposed directly at the embedded boundary through the
analytical reconstruction formula derived in Secs.~\ref{sec:bilinear} and
\ref{sec:trilinear}. As shown in
Fig.~\ref{fig:comparison_reconstruction_methods}(b), this permits a compact
nearest-neighbour Cartesian stencil. However, with the classical ghost-cell
placement, this compact stencil does not necessarily satisfy the stability
criterion for every boundary configuration. The CGC\_ES formulation introduced
in Sec.~\ref{sec:extended_reconstruction} addresses this by enlarging the
stencil where required while retaining the same classical ghost-cell
placement.

HGC instead modifies the admissible ghost-cell placement near the embedded
boundary. By allowing ghost-cell centers on either side of the interface, the
stability criterion can be satisfied while retaining the nearest-neighbour
stencil, as illustrated in
Fig.~\ref{fig:comparison_reconstruction_methods}(c). 

\begin{table}[htp]
\centering
\caption{Comparison of ghost-cell reconstruction strategies.}
\resizebox{\textwidth}{!}{
\begin{tabular}{lccccc}
\hline
Method &
Stable &
Compact stencil &
Iterative &
Weight storage &
Direct reconstruction / EB gradients \\
\hline
Image-point IBM \cite{MittalDBNVL:2008} &
-- &
No &
Yes &
Yes &
No \\
CGC &
No &
Yes &
No &
No &
Yes \\
CGC\_ES &
Yes &
No &
No &
No &
Yes \\
Present HGC &
Yes &
Yes &
No &
No &
Yes \\
\hline
\end{tabular}
}
\label{tab:comparison}
\end{table}

The principal characteristics are summarized in
Table~\ref{tab:comparison}. Taken together, the formulations separate the
effects of reconstruction procedure, ghost-cell placement, satisfaction of
the stability criterion, and stencil compactness. CGC isolates the effect of
replacing image-point reconstruction with direct analytical reconstruction,
CGC\_ES isolates the effect of satisfying the stability criterion through
stencil enlargement, and HGC demonstrates that the same stability requirement
can be satisfied while retaining the compact nearest-neighbour stencil.
}

\section{Analytical ghost-cell reconstruction in two dimensions}
\label{sec:bilinear}
\begin{figure}
 \centering
 \begin{minipage}{\linewidth}
 \centering
 \includegraphics[width=2.5in,trim={0.6cm 0.6cm 2cm 2.cm},clip]{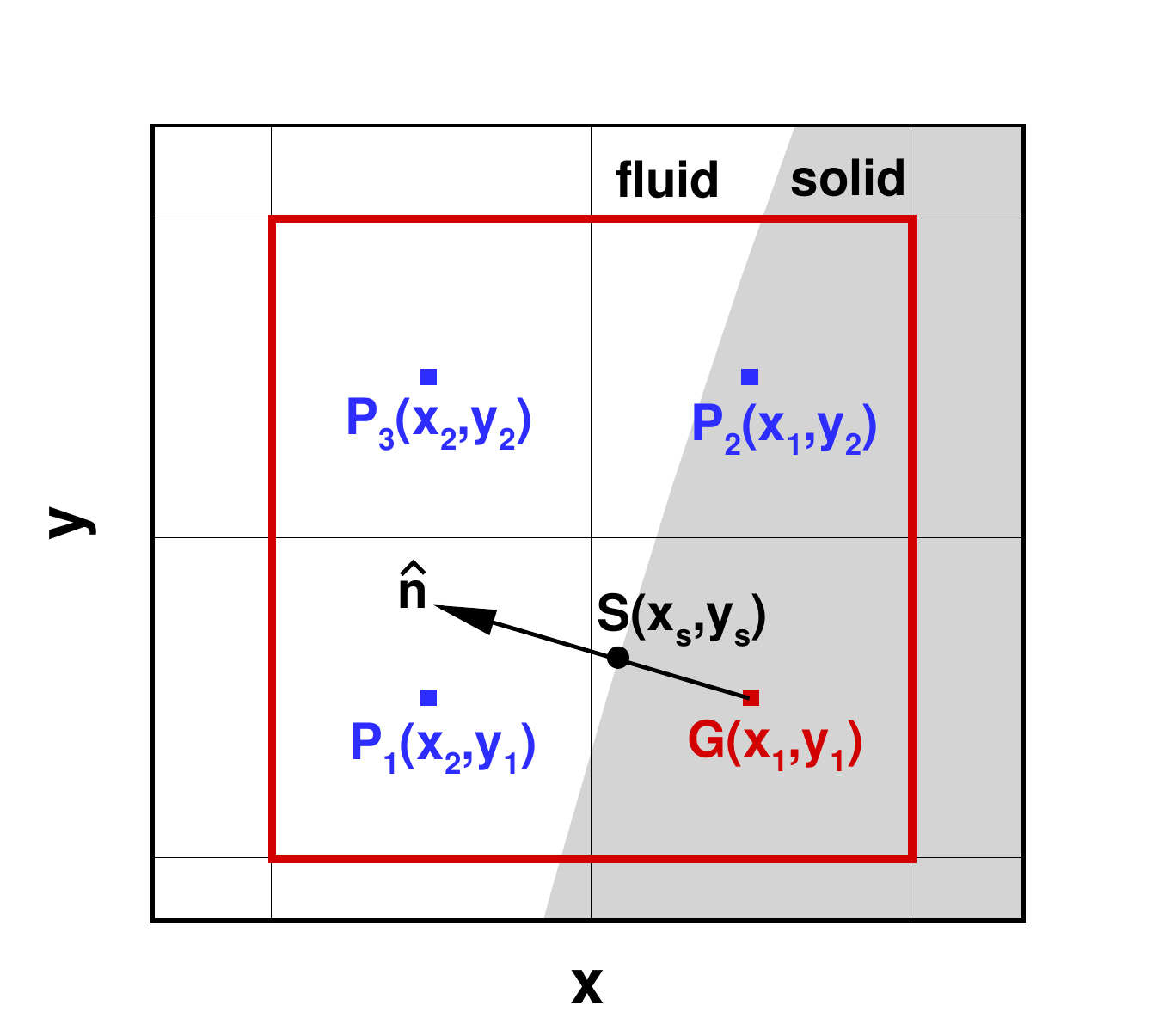} 
 \end{minipage} 
 \caption{Stencil used in the bilinear interpolation for a ghost cell $G$: $P_1, P_2, P_3$ are the neighbour cells
 and $S$ is the intersection of the EB with the surface normal ($\hat{n}$) passing through $G$. 
 }
 \label{fig:stencil_2d}
\end{figure}
Assume a bilinear solution in two dimensions of the form
\begin{equation*}
\phi(x,y)
=
C_0 + C_1 x + C_2 y + C_3 xy ,
\end{equation*}
which provides second-order spatial accuracy. The unknown coefficients are
determined using the solution values at three neighboring cells of the ghost
cell together with a boundary condition imposed on the embedded boundary.

As shown in Fig.~\ref{fig:stencil_2d}, let $G(x_1,y_1)$ denote the ghost cell.
The three neighboring Cartesian cells and the boundary-intersection point are
\begin{equation*}
P_1=(x_2,y_1), \qquad
P_2=(x_1,y_2), \qquad
P_3=(x_2,y_2), \qquad
S=(x_s,y_s),
\end{equation*}
where $S$ is the intersection of the embedded boundary with the surface normal
passing through $G$. Let the unit surface normal pointing into the fluid region
be
\begin{equation*}
\hat{\boldsymbol{n}}=(n_x,n_y),
\end{equation*}
and define the sign vector as
\begin{equation*}
\boldsymbol{e}
=
(e_x,e_y)
=
\left(\sgn(n_x),\sgn(n_y)\right).
\end{equation*}
Depending on the orientation of the surface normal, the diagonally opposite
stencil point satisfies
\begin{equation*}
(x_2,y_2)
=
(x_1,y_1)
+
\left(
\sgn(n_x)\Delta x,\,
\sgn(n_y)\Delta y
\right),
\end{equation*}
where $\Delta x$ and $\Delta y$ are the local Cartesian mesh spacings at the
ghost cell $G$ in the $x$- and $y$-directions, respectively.

\subsection{Boundary Conditions Settings}
The ghost-cell value $\phi_G$ is reconstructed for either Dirichlet or Neumann
boundary conditions imposed at $S(x_s,y_s)$ on the embedded boundary. The
derivations of the analytical expressions for both boundary-condition types
are provided in \ref{sec:proof2d}.
\subsubsection{Dirichlet type}
\begin{eqnarray}
\phi_G &=& \omega_1\phi_1+ \omega_2\phi_2 -\omega_1\omega_2\phi_3 +(1-\omega_1)(1-\omega_2) \phi_s \label{eqn:phi_G_Dir_2d}
\end{eqnarray}
where,
\begin{eqnarray*}
 \omega_1 = \frac{x_1-x_s}{x_2 - x_s},\quad
 \omega_2 =\frac{y_1-y_s} {y_2 - y_s} 
\end{eqnarray*}
\subsubsection{Neumann type}
\begin{eqnarray}
\phi_G = \omega_1\phi_1+ \omega_2\phi_2 +(1-\omega_1-\omega_2)\phi_3 +\omega_s \bdelux{\phi}{n}
\label{eqn:phi_G_Neu_2d}
\end{eqnarray}
where, for $a_{1} = n_x(y_1-y_s)$, $a_{2} = n_x(y_2-y_s),
 b_{1} = n_y(x_1-x_s),$ and $b_{2} = n_y(x_2-x_s)$ we set
\begin{equation*}
 \omega_1 =\frac{(a_{2} + b_{1})}{(a_{2}+b_{2})}, \quad
 \omega_2 = \frac{(a_{1} + b_{2})}{(a_{2}+b_{2})}, \quad
 \omega_s = -\frac{(x_2-x_1)(y_2-y_1)}{(a_{2}+b_{2})}.
\end{equation*}
\\Note that if $n_x=0$, then $x_s=x_1, \omega_1=0, \omega_2=1$ and if $n_y=0$, then $y_s=y_1, \omega_1=1, \omega_2=0$. 

\section{Analytical ghost-cell reconstruction in three dimensions}
\label{sec:trilinear}
Assume a trilinear solution in three dimensions of the form
\begin{equation*}
\phi(x,y,z)
=
C_0 + C_1 x + C_2 y + C_3 z
+ C_4 xy + C_5 yz + C_6 xz + C_7 xyz ,
\end{equation*}
which provides second-order spatial accuracy. The unknown coefficients are
determined using the solution values at seven neighboring cells of the ghost
cell together with a boundary condition imposed on the embedded boundary.

Let $G(x_1,y_1,z_1)$ denote the ghost cell. The seven neighboring Cartesian
cells and the boundary-intersection point are
\begin{equation*}
\begin{aligned}
P_1 &= (x_2,y_1,z_1), \qquad
P_2 = (x_1,y_2,z_1), \qquad
P_3 = (x_1,y_1,z_2), \qquad
P_4 = (x_2,y_2,z_1),\\
P_5 &= (x_1,y_2,z_2), \qquad
P_6 = (x_2,y_1,z_2), \qquad
P_7 = (x_2,y_2,z_2), \qquad
S = (x_s,y_s,z_s).
\end{aligned}
\end{equation*}
where $S$ is the intersection of the embedded boundary with the surface normal
passing through $G$. Let the unit surface normal pointing into the fluid region
be
\begin{equation*}
\hat{\boldsymbol{n}}=(n_x,n_y,n_z),
\end{equation*}
and define the sign vector as
\begin{equation*}
\boldsymbol{e}
=
(e_x,e_y,e_z)
=
\left(\sgn(n_x),\sgn(n_y),\sgn(n_z)\right).
\end{equation*}
Depending on the orientation of the surface normal, the diagonally opposite
stencil point satisfies
\begin{equation*}
(x_2,y_2,z_2)
=
(x_1,y_1,z_1)
+
\left(
e_x\Delta x,\,
e_y\Delta y,\,
e_z\Delta z
\right),
\end{equation*}
where $\Delta x$, $\Delta y$, and $\Delta z$ are the local Cartesian mesh
spacings at the ghost cell $G$ in the $x$-, $y$-, and $z$-directions,
respectively.

\subsection{Boundary Conditions Settings}
The ghost-cell value $\phi_G$ is reconstructed for either Dirichlet or Neumann
boundary conditions imposed at $S(x_s,y_s,z_s)$ on the embedded boundary. The
derivations of the analytical expressions for both boundary-condition types
are provided in \ref{sec:proof3d}.
\subsubsection{Dirichlet type}
\begin{eqnarray}
\phi_G &=& \omega_1\phi_1+ \omega_2\phi_2+ \omega_3\phi_3 -\omega_1\omega_2\phi_4 -\omega_2\omega_3\phi_5 -\omega_1\omega_3\phi_6+\omega_1\omega_2\omega_3\phi_7 + (1-\omega_1)(1-\omega_2)(1-\omega_3) \phi_s \label{eqn:phi_G_Dir_3d}
\end{eqnarray}
where,
\begin{eqnarray*}
 \omega_1 &=& \frac{x_1-x_s}{x_2 - x_s}, \quad
 \omega_2 =\frac{y_1-y_s} {y_2 - y_s}, \quad 
 \omega_3 =\frac{z_1-z_s}{z_2 - z_s}.
\end{eqnarray*}
\subsubsection{Neumann type}
\begin{equation}
\phi_G
=
\omega_1\phi_1
+\omega_2\phi_2
+\omega_3\phi_3
+\omega_4\phi_4
+\omega_5\phi_5
+\omega_6\phi_6
+\omega_7\phi_7
+\omega_s \frac{\partial \phi}{\partial n},
\label{eqn:phi_G_Neu_3d}
\end{equation}
where
\begin{equation*}
\begin{aligned}
\omega_1 &=
\frac{a_{22}+b_{12}+c_{12}}
     {a_{22}+b_{22}+c_{22}},
\qquad
\omega_2 =
\frac{a_{12}+b_{22}+c_{21}}
     {a_{22}+b_{22}+c_{22}},
\qquad
\omega_3 =
\frac{a_{21}+b_{21}+c_{22}}
     {a_{22}+b_{22}+c_{22}},
\\[1mm]
\omega_4 &=
-\frac{a_{12}+b_{12}+c_{11}}
      {a_{22}+b_{22}+c_{22}},
\qquad
\omega_5 =
-\frac{a_{11}+b_{21}+c_{21}}
      {a_{22}+b_{22}+c_{22}},
\qquad
\omega_6 =
-\frac{a_{21}+b_{11}+c_{12}}
      {a_{22}+b_{22}+c_{22}},
\\[1mm]
\omega_7 &=
\frac{a_{11}+b_{11}+c_{11}}
     {a_{22}+b_{22}+c_{22}},
\qquad
\omega_s =
-\frac{(x_2-x_1)(y_2-y_1)(z_2-z_1)}
      {a_{22}+b_{22}+c_{22}}.
\end{aligned}
\end{equation*}
with
\begin{equation*}
\begin{aligned}
a_{ij} &= n_x(y_i-y_s)(z_j-z_s),\\
b_{ij} &= n_y(x_i-x_s)(z_j-z_s),\\
c_{ij} &= n_z(x_i-x_s)(y_j-y_s),
\qquad i,j=1,2.
\end{aligned}
\end{equation*}
The denominator $a_{22}+b_{22}+c_{22}$ remains nonzero provided that
$x_2\neq x_s$, $y_2\neq y_s$, and $z_2\neq z_s$. These conditions are
inherently satisfied by the ghost cells defined in Sec.~\ref{sec:sort}.
Furthermore, if $n_x=0$, then $x_s=x_1$; similarly, $n_y=0$ implies
$y_s=y_1$, and $n_z=0$ implies $z_s=z_1$. The reconstruction weights are
independent of the sense of the surface-normal vector.
\section{Analytical evaluation of surface quantities on the embedded boundary}
\label{sec:surface_quantities}

The same bilinear and trilinear reconstruction relations used for ghost-cell
reconstruction can also be rearranged to evaluate the solution and its spatial
derivatives directly on the embedded boundary. The Dirichlet reconstruction
relations in Eqs.~\eqref{eqn:phi_G_Dir_2d} and
\eqref{eqn:phi_G_Dir_3d} can be rearranged to obtain the boundary value
$\phi_s$ from the ghost-cell value and the neighboring Cartesian values.

Similarly, rearranging the Neumann reconstruction relations in
Eqs.~\eqref{eqn:phi_G_Neu_2d} and \eqref{eqn:phi_G_Neu_3d} yields the normal
derivative at the boundary-intersection point $S$. Expressing this derivative
as $\nabla\phi\cdot\hat{\boldsymbol{n}}$ and separating the contributions
associated with the Cartesian components of
$\hat{\boldsymbol{n}}$ gives analytical expressions for the individual
spatial derivatives on the embedded boundary.

\subsection{Two-dimensional formulation}
In two dimensions, Eq.~\eqref{eqn:phi_G_Neu_2d} can be rearranged to obtain
the normal derivative directly as
\begin{eqnarray}
\left.\bdelux{\phi}{n}\right|_s &=& \frac{(a_{2} + b_{1})\phi_1+ (a_{1} + b_{2})\phi_2 -(a_{1} + b_{1})\phi_3 -(a_{2}+b_{2})\phi_G} {(x_2-x_1)(y_2-y_1)}, \nonumber\\
\left.\nabla \phi \cdot \hat{n}\right|_s  &=& \frac{(n_x y_{2} + n_y x_{1})\phi_1+ (n_x y_{1} + n_y x_{2})\phi_2 -(n_x y_{1} + n_y x_{1})\phi_3 -(n_x y_{2}+n_y x_{2})\phi_G} {(x_2-x_1)(y_2-y_1)},\nonumber\\
\left. n_x \bdelux{\phi}{x}\right|_s + \left. n_y \bdelux{\phi}{y}\right|_s  &=& \frac{ n_x (y_{2} \phi_1 +  y_{1} \phi_2 -  y_{1} \phi_3 - y_{2}\phi_G) + n_y (x_{1} \phi_1 +  x_{2}\phi_2 -  x_{1}\phi_3    -  x_{2}\phi_G) } {(x_2-x_1)(y_2-y_1)}.
\end{eqnarray}
By comparison of the terms with $n_x$ and $n_y$ on both sides of the above equation, we have,
\begin{eqnarray}
\left. \bdelux{\phi}{x}\right|_s &=&  \left(\frac{y_2-y_s}{y_2-y_1}\right) \left[ \frac{\phi_1-\phi_G}{x_2-x_1}\right] + \left(\frac{y_s-y_1}{y_2-y_1}\right) \left[ \frac{\phi_3-\phi_2}{x_2-x_1}\right] \nonumber\\
\left. \bdelux{\phi}{y}\right|_s &=&  \left(\frac{x_2-x_s}{x_2-x_1}\right) \left[ \frac{\phi_2-\phi_G}{y_2-y_1}\right] + \left(\frac{x_s-x_1}{x_2-x_1}\right) \left[ \frac{\phi_3-\phi_1}{y_2-y_1}\right] \label{eqn:derivative_bilinear}
\end{eqnarray}
Above equations are essentially linear interpolation of $\delux{\phi}{x}(y)$ in $y-$direction and $\delux{\phi}{y}(x)$ in $x-$direction because $\delux{\phi}{x}$, $\delux{\phi}{y}$ depend only on $y, x$, respectively, in bilinear approximation.

\subsection{Three-dimensional formulation}

In three dimensions, Eq.~\eqref{eqn:phi_G_Neu_3d} can be rearranged to obtain
the normal derivative directly as
\begin{equation}
\begin{aligned}
\left.\frac{\partial \phi}{\partial n}\right|_s
=
\frac{1}{(x_2-x_1)(y_2-y_1)(z_2-z_1)}
\Big[
& (a_{22}+b_{12}+c_{12})\phi_1
 +(a_{12}+b_{22}+c_{21})\phi_2 \\
& +(a_{21}+b_{21}+c_{22})\phi_3
 -(a_{12}+b_{12}+c_{11})\phi_4 \\
& -(a_{11}+b_{21}+c_{21})\phi_5
 -(a_{21}+b_{11}+c_{12})\phi_6 \\
& +(a_{11}+b_{11}+c_{11})\phi_7
 -(a_{22}+b_{22}+c_{22})\phi_G
\Big].
\end{aligned}
\label{eqn:normal_derivative_trilinear}
\end{equation}

Since $a_{ij}$, $b_{ij}$, and $c_{ij}$ are proportional exclusively to
$n_x$, $n_y$, and $n_z$, respectively, the terms in
Eq.~\eqref{eqn:normal_derivative_trilinear} can be separated according to the
components of the surface-normal vector. Using
\begin{equation*}
\left.\frac{\partial \phi}{\partial n}\right|_s
=
\left.
\left(
n_x\frac{\partial\phi}{\partial x}
+
n_y\frac{\partial\phi}{\partial y}
+
n_z\frac{\partial\phi}{\partial z}
\right)\right|_s ,
\end{equation*}
and comparing the coefficients of $n_x$, $n_y$, and $n_z$ gives the Cartesian
derivatives directly on the embedded boundary:
\begin{align}
\left.\frac{\partial\phi}{\partial x}\right|_s
={}&
\left(\frac{y_2-y_s}{y_2-y_1}\right)
\left(\frac{z_2-z_s}{z_2-z_1}\right)
\left[\frac{\phi_1-\phi_G}{x_2-x_1}\right]
+
\left(\frac{y_s-y_1}{y_2-y_1}\right)
\left(\frac{z_2-z_s}{z_2-z_1}\right)
\left[\frac{\phi_4-\phi_2}{x_2-x_1}\right]
\nonumber\\
&+
\left(\frac{y_2-y_s}{y_2-y_1}\right)
\left(\frac{z_s-z_1}{z_2-z_1}\right)
\left[\frac{\phi_6-\phi_3}{x_2-x_1}\right]
+
\left(\frac{y_s-y_1}{y_2-y_1}\right)
\left(\frac{z_s-z_1}{z_2-z_1}\right)
\left[\frac{\phi_7-\phi_5}{x_2-x_1}\right],
\label{eqn:derivative_trilinear_x}
\\[1ex]
\left.\frac{\partial\phi}{\partial y}\right|_s
={}&
\left(\frac{x_2-x_s}{x_2-x_1}\right)
\left(\frac{z_2-z_s}{z_2-z_1}\right)
\left[\frac{\phi_2-\phi_G}{y_2-y_1}\right]
+
\left(\frac{x_s-x_1}{x_2-x_1}\right)
\left(\frac{z_2-z_s}{z_2-z_1}\right)
\left[\frac{\phi_4-\phi_1}{y_2-y_1}\right]
\nonumber\\
&+
\left(\frac{x_2-x_s}{x_2-x_1}\right)
\left(\frac{z_s-z_1}{z_2-z_1}\right)
\left[\frac{\phi_5-\phi_3}{y_2-y_1}\right]
+
\left(\frac{x_s-x_1}{x_2-x_1}\right)
\left(\frac{z_s-z_1}{z_2-z_1}\right)
\left[\frac{\phi_7-\phi_6}{y_2-y_1}\right],
\label{eqn:derivative_trilinear_y}
\\[1ex]
\left.\frac{\partial\phi}{\partial z}\right|_s
={}&
\left(\frac{x_2-x_s}{x_2-x_1}\right)
\left(\frac{y_2-y_s}{y_2-y_1}\right)
\left[\frac{\phi_3-\phi_G}{z_2-z_1}\right]
+
\left(\frac{x_s-x_1}{x_2-x_1}\right)
\left(\frac{y_2-y_s}{y_2-y_1}\right)
\left[\frac{\phi_6-\phi_1}{z_2-z_1}\right]
\nonumber\\
&+
\left(\frac{x_2-x_s}{x_2-x_1}\right)
\left(\frac{y_s-y_1}{y_2-y_1}\right)
\left[\frac{\phi_5-\phi_2}{z_2-z_1}\right]
+
\left(\frac{x_s-x_1}{x_2-x_1}\right)
\left(\frac{y_s-y_1}{y_2-y_1}\right)
\left[\frac{\phi_7-\phi_4}{z_2-z_1}\right].
\label{eqn:derivative_trilinear_z}
\end{align}

These expressions may also be interpreted as bilinear interpolations, on the
corresponding coordinate planes, of directional differences across the
trilinear stencil. Consequently, the three Cartesian derivatives at $S$ are
evaluated from the ghost-cell value and the neighboring Cartesian cell values,
without solving explicitly for the polynomial coefficients $C_j$.

The analytically evaluated boundary values and gradients can be used directly
to compute pressure forces, wall stresses, drag, and lift, without introducing
a separate surface-interpolation or reconstruction procedure. No filtering is
applied to the surface quantities or force histories reported in
Secs.~\ref{sec:cylinder} and \ref{sec:airfoil}; nevertheless, the
stability-preserving reconstructions produce smooth surface distributions and
force histories.

\section{Extended-stencil reconstruction}
\label{sec:extended_reconstruction}
\subsection{Additional ghost-cell layers}
Additional layers of ghost cells may be required in applications involving,
for example, moving boundaries that expose fresh fluid cells, mixed
derivatives, flux-limited discretizations, higher-order spatial schemes, or
post-processing operations. When a second ghost-cell layer is required, solid
cells satisfying
\begin{equation*}
\psi \leq 1.5\Delta n
\end{equation*}
are included in the ghost-cell set.

The analytical reconstruction described in
Secs.~\ref{sec:bilinear} and \ref{sec:trilinear} is retained, but the Cartesian
stencil spacing may be increased independently in each coordinate direction.
For a ghost cell centered at
$\boldsymbol{x}_1=(x_1,y_1,z_1)$, let
$\boldsymbol{x}_s=(x_s,y_s,z_s)$ denote the corresponding
boundary-intersection point. The diagonally opposite stencil point is selected
as
\begin{equation}
\boldsymbol{x}_2
=
\boldsymbol{x}_1
+
(r_1e_x\Delta x,\,
 r_2e_y\Delta y,\,
 r_3e_z\Delta z),
\qquad
r_i
=
\max\left\{
1,\,
\left\lceil
\frac{2|n_i\psi|}{\Delta x_i}
\right\rceil
\right\}
=
\max\left\{
1,\,
\left\lceil
\frac{2|x_{1,i}-x_{s,i}|}{\Delta x_i}
\right\rceil
\right\}.
\label{eqn:stencil_anisotropic}
\end{equation}
where $\lceil q\rceil$ denotes the ceiling function, i.e., the smallest
integer greater than or equal to $q$. The integer $r_i$ is the stencil-spacing
multiplier in coordinate direction $i$, and the maximum with unity ensures
that $r_i\geq 1$, including when the corresponding component $n_i$ of the
boundary-normal vector vanishes. Here, $x_{1,i}$ and $x_{s,i}$ are the
$i$th components of $\boldsymbol{x}_1$ and $\boldsymbol{x}_s$, respectively,
and $\Delta x_i$ is the local mesh spacing in direction $i$.

By definition of $r_i$,
\begin{equation*}
\frac{|x_{1,i}-x_{s,i}|}{r_i\Delta x_i}
\leq \frac{1}{2},
\end{equation*}
so that the normalized reconstruction distance is bounded by $0.5$ in each
coordinate direction. Since $\alpha_{\max}=0.5$ was found to be a sufficient
stability bound for all spatial and temporal discretizations examined in
\cite{RapakaS:2018}, the extended stencil is constructed to satisfy the
corresponding reconstruction-stability constraint. The applicability of this
criterion to the nonlinear incompressible Navier--Stokes simulations
considered here is examined in Sec.~\ref{sec:stability_results}.

The stencil multiplier $r_i$ increases only as needed to satisfy the
normalized-distance constraint. Consequently, ghost cells near the embedded
boundary may retain $r_i=1$ in one or more coordinate directions, whereas
cells farther inside the solid region generally require a wider stencil.


For anisotropic grids, Eq.~\eqref{eqn:stencil_anisotropic} can similarly be
applied independently in each Cartesian direction using the corresponding
local mesh spacing $\Delta x_i$. The resulting stencil selection is consistent
with the same normalized-distance constraint described above. 
\subsection{Classical ghost cells with an extended stencil (CGC\_ES)}

The extended-stencil construction described above can also be applied to the
first layer of classical ghost cells. The resulting formulation, denoted
CGC\_ES (Classical Ghost Cells with Extended Stencil), retains the traditional
placement of ghost-cell centers strictly inside the solid region but enlarges
the reconstruction stencil according to
Eq.~\eqref{eqn:stencil_anisotropic} wherever necessary to satisfy the
stability criterion in Eq.~\eqref{eqn:stability_alpha_max}. In the numerical
assessment of Sec.~\ref{sec:stability_results}, CGC\_ES therefore serves as a
stability-preserving reference based on classical ghost-cell placement.


\narsimha{
\section{Stability of hybrid ghost cells (HGC) versus classical ghost cells (CGC)}\label{sec:stability_results}
The stability criterion used in the present comparisons follows from the
linear analysis of ghost-cell reconstruction presented in
\cite{RapakaS:2018}. In that study, reconstruction stability was examined
using $\varepsilon$-pseudospectra for scalar advection in one and two
dimensions with several spatial and temporal discretization schemes. A
sufficient condition for linear stability in two dimensions was found to be
\begin{equation}
    \alpha_x,\alpha_y \leq \alpha_{\max},
    \label{eqn:stability_alpha_max}
\end{equation}
where
\begin{equation*}
    \alpha_x \coloneqq
    \frac{x_s-x_1}{x_2-x_1},
    \qquad
    \alpha_y \coloneqq
    \frac{y_s-y_1}{y_2-y_1},
\end{equation*}
represent the projected ghost-cell-to-boundary distances normalized by the
corresponding reconstruction-stencil spacing. The limiting value
$\alpha_{\max}$ depends on the spatial and temporal discretization. A value
of $\alpha_{\max}=0.5$ was sufficient for all schemes examined in
\cite{RapakaS:2018}. For the RK3-CDS discretization used in the present
incompressible Navier--Stokes solver \cite{RapakaS:2016},
$\alpha_{\max}=0.77$ at a maximum Courant number
$C=u\Delta t/\Delta x=1.7$.

The objective of the present section is to examine whether this linear
stability criterion remains a useful indicator for nonlinear incompressible
Navier--Stokes simulations.
The incompressible Navier--Stokes solver and immersed-boundary framework employed here were previously developed, verified, and validated in \cite{RapakaS:2016}. The governing equations, spatial discretization, time-integration scheme, and pressure-projection methodology remain unchanged. The only modification introduced in the present study is the ghost-cell reconstruction procedure.

Three reconstruction strategies are considered within the same solver framework:

\begin{enumerate}
    \item Classical ghost cells (CGC), corresponding to the conventional formulation in which ghost-cell centers lie strictly inside the solid region and reconstruction is performed using the nearest-neighbour stencil.
    
    \item Hybrid ghost cells (HGC), introduced by \cite{RapakaS:2018}, whose centers may lie on either side of the embedded boundary and therefore remain closer to the interface.
    
    \item Classical ghost cells with extended stencil (CGC\_ES), introduced in Sec.~\ref{sec:extended_reconstruction}, which retain the classical ghost-cell placement but enlarge the reconstruction stencil whenever necessary to satisfy the stability criterion Eq.~(\ref{eqn:stability_alpha_max}).
\end{enumerate}

CGC and HGC employ the nearest-neighbour reconstruction stencil
($r_i=1$), whereas CGC\_ES uses the extended stencil defined by
Eq.~\eqref{eqn:stencil_anisotropic}. Comparing CGC with CGC\_ES assesses the
effect of satisfying the stability criterion while retaining the classical
ghost-cell placement. Comparing CGC\_ES with HGC contrasts two
stability-preserving strategies and assesses the benefit of satisfying the
criterion with a compact nearest-neighbour stencil.

}



\subsection{Flow past a circular cylinder}\label{sec:cylinder}
Flow past a circular cylinder of diameter $d$ is simulated at $Re=100$ using
the CGC, HGC, and CGC\_ES reconstruction strategies. The computational domain
has dimensions $35d\times30d\times0.4d$ and is discretized using a
$384\times384\times4$ Cartesian grid. The mesh is locally uniform in the
vicinity of the cylinder, with a spacing of $0.015d$, and is gradually
stretched toward the outer boundaries with a stretching ratio of $3\%$.
A uniform velocity $U_\infty$ is prescribed at the inflow, and a parabolic
sponge layer is applied near the outflow to reduce spurious reflections.

The simulations are performed at a Courant number of $C=1.7$, corresponding
to the upper stability limit of the RK3-CDS scheme in the absence of an
embedded boundary \citep{RapakaS:2018}. For CGC, HGC, and CGC\_ES,
respectively, the maximum normalized reconstruction distances are
\begin{align*}
\max(\alpha_x) &= 0.924,\;0.464,\;0.481,\\
\max(\alpha_y) &= 0.926,\;0.465,\;0.485.
\end{align*}
Thus, the CGC reconstruction violates the stability criterion in
Eq.~\eqref{eqn:stability_alpha_max}, since both
$\max(\alpha_x)$ and $\max(\alpha_y)$ exceed
$\alpha_{\max}=0.77$. In contrast, HGC and CGC\_ES satisfy the criterion in
both coordinate directions.
\begin{figure}[htp]
 \centering
 \begin{tabular}{cc}
   \begin{minipage}{0.49\linewidth}
     \centering
     \includegraphics[width=2.75in,trim={0.in 0.in 0.in 0.in}]{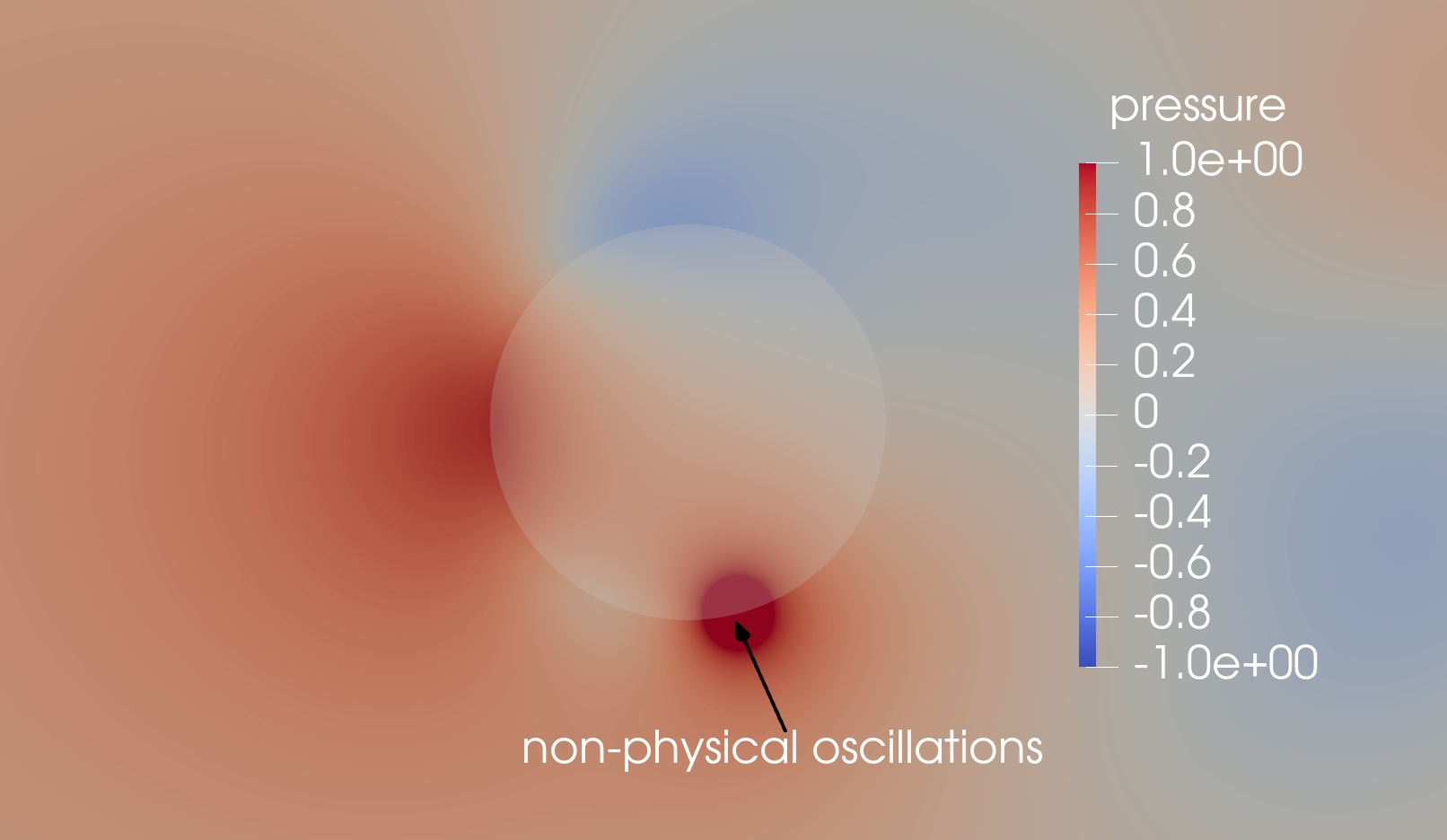}\\
     {\scriptsize(a) Pressure with CGC}
   \end{minipage} 
   \begin{minipage}{0.49\linewidth}
     \centering
     \includegraphics[width=2.75in,trim={0.in 0.in 0.in 0.in}]{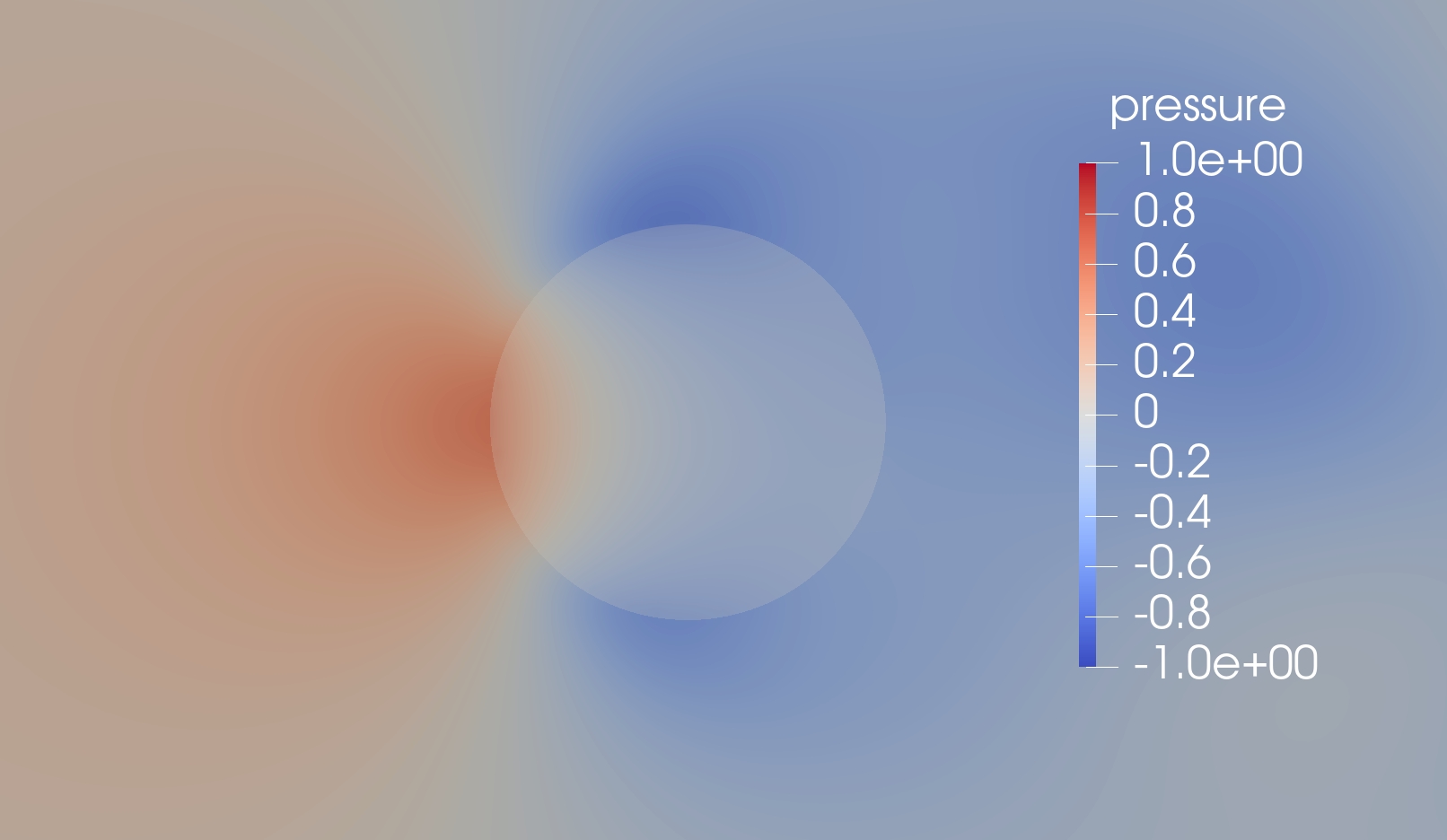}\\
     {\scriptsize(b) Pressure with HGC}
   \end{minipage} \\
   \begin{minipage}{0.49\linewidth}
     \centering
     \includegraphics[width=2.8in,trim={0.in 0.in 0.in 0.in}]{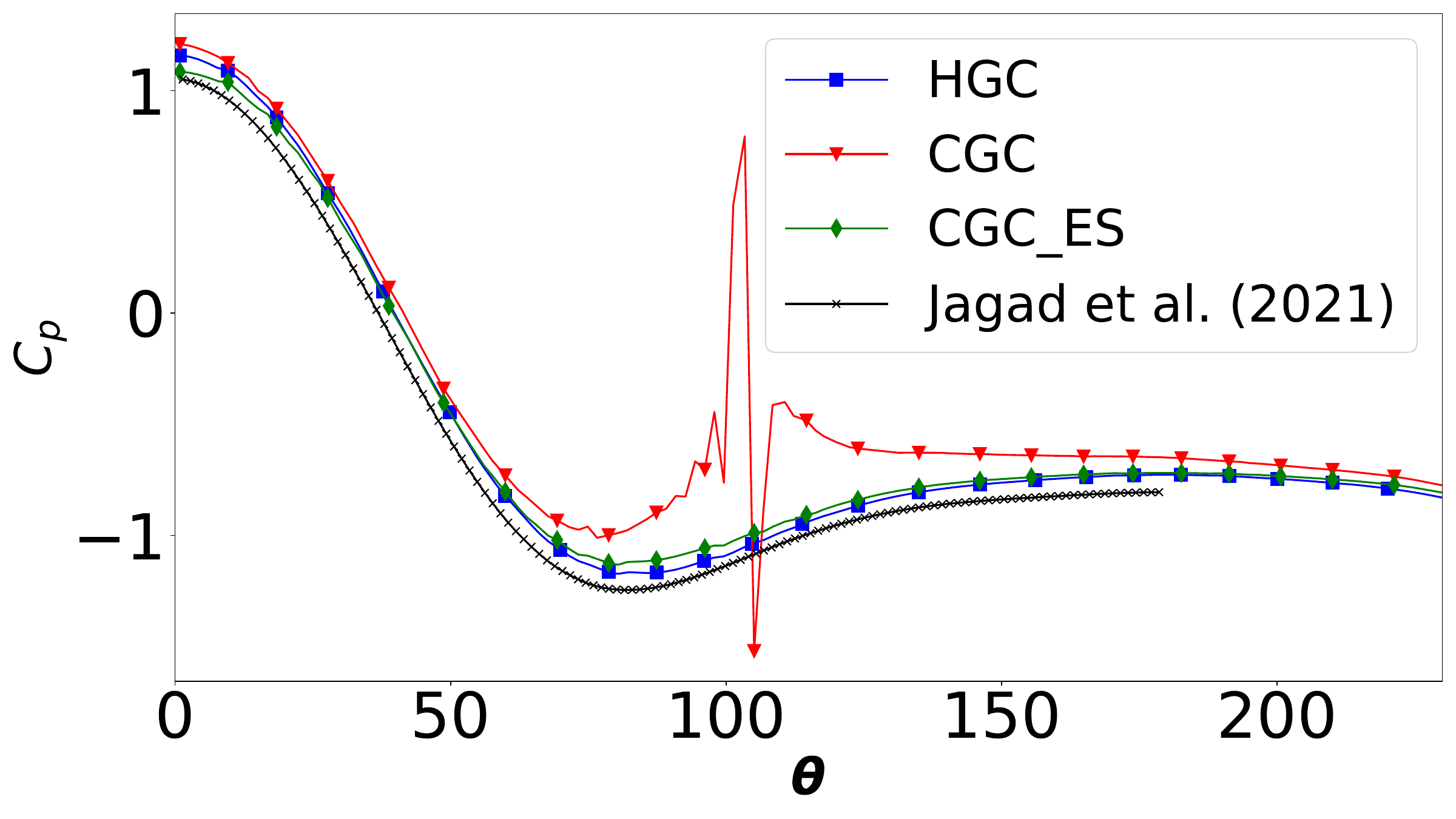}\\
     {\scriptsize(c) Distribution of time-averaged $C_p$}
   \end{minipage} 
   \begin{minipage}{0.49\linewidth}
     \centering
     \includegraphics[width=2.8in,trim={0.in 0.in 0.in 0.in}]{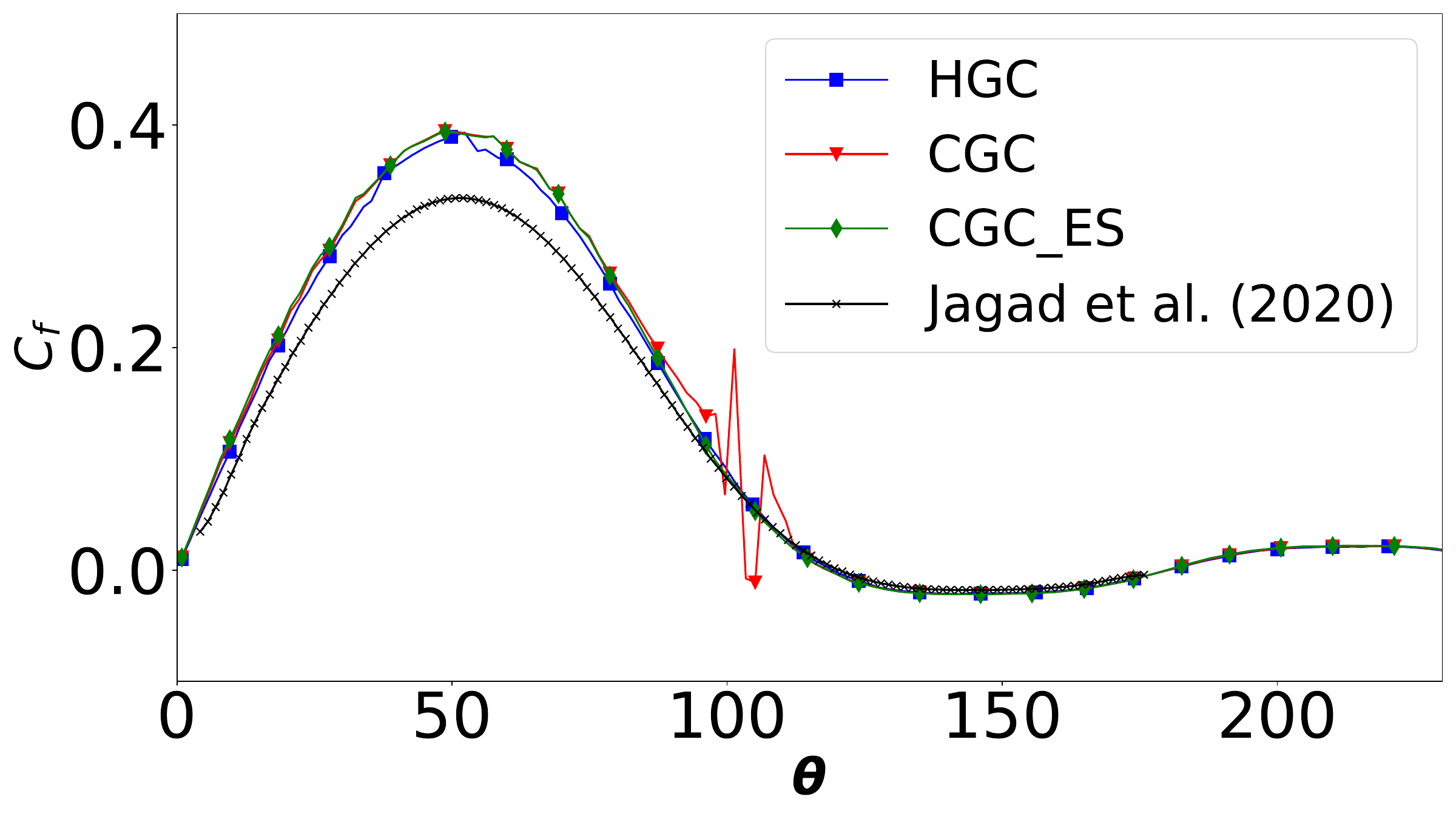}\\
     {\scriptsize(d) Distribution of time-averaged $C_f$}
   \end{minipage} \\
   \begin{minipage}{0.49\linewidth}
     \centering
     \includegraphics[width=2.7in,trim={0.in 0.in 0.in 0.in}]{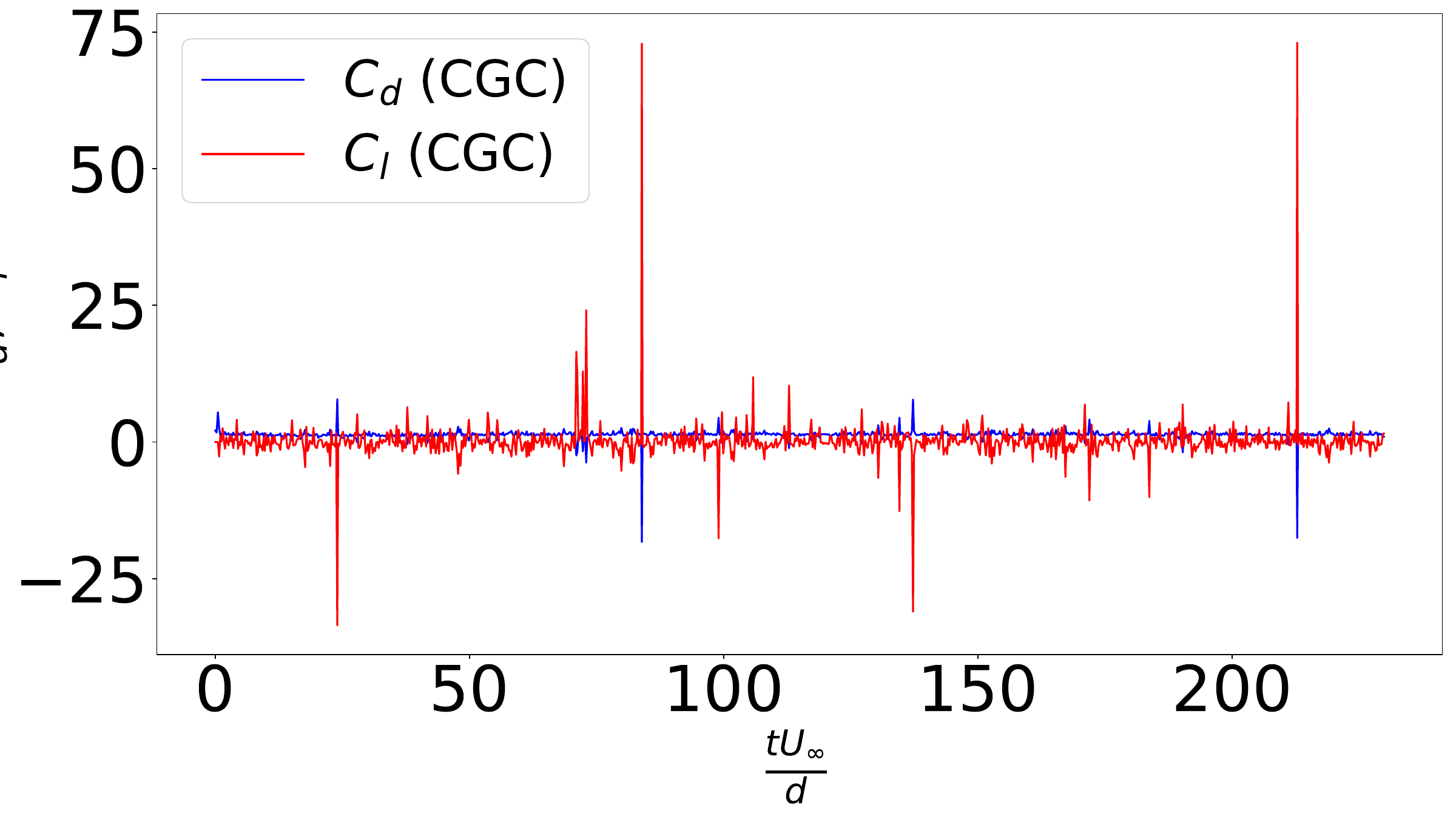}\\
     {\scriptsize(e) Time evolution of $C_d, C_l$ (CGC)}
   \end{minipage} 
   \begin{minipage}{0.49\linewidth}
     \centering
     \includegraphics[width=2.8in,trim={0.in 0.in 0.in 0.in}]{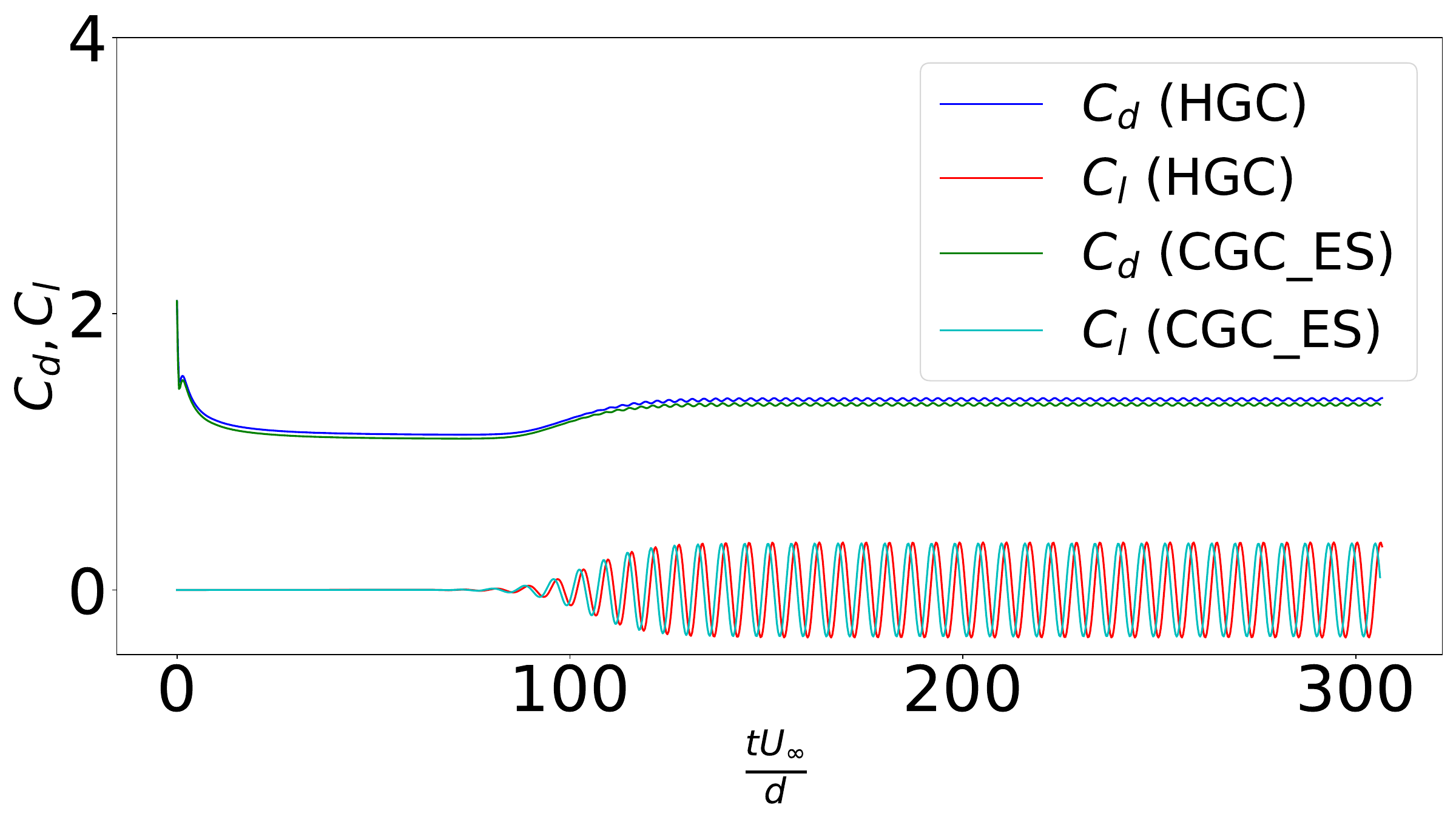}\\
     {\scriptsize(f) Time evolution of $C_d, C_l$ (HGC, CGC\_ES)}
   \end{minipage}
 \end{tabular}
\caption{
Flow past a circular cylinder at $Re=100$.
(a) Pressure field at $t=229.8\,d/U_\infty$ obtained with CGC;
(b) pressure field at the same instant obtained with HGC;
(c) time-averaged pressure coefficient $C_p$ for CGC, HGC, and CGC\_ES;
(d) time-averaged skin-friction coefficient $C_f$ for CGC, HGC, and
CGC\_ES;
(e) time histories of the drag and lift coefficients for CGC;
(f) time histories of the drag and lift coefficients for HGC and CGC\_ES.
CGC violates the stability criterion in
Eq.~\eqref{eqn:stability_alpha_max} and exhibits spurious oscillations,
whereas HGC and CGC\_ES satisfy the criterion and produce stable solutions.
HGC achieves this while retaining the nearest-neighbour reconstruction
stencil. The angle $\theta$ is measured counter-clockwise from the front
stagnation point.
}
 \label{fig:compare_CGC_HGC}
\end{figure}

Figures~\ref{fig:compare_CGC_HGC}(a) and
\ref{fig:compare_CGC_HGC}(b) show the pressure fields obtained with CGC and
HGC, respectively. The CGC solution develops spurious oscillations near
$\theta\approx100^\circ$, where the reconstruction violates the stability
criterion in Eq.~\eqref{eqn:stability_alpha_max}. In contrast, the HGC
solution remains smooth because its reconstruction satisfies the stability
criterion.

The time-averaged pressure and skin-friction coefficients shown in
Figs.~\ref{fig:compare_CGC_HGC}(c) and
\ref{fig:compare_CGC_HGC}(d) further illustrate this behavior. CGC produces
non-physical oscillations near $\theta\approx100^\circ$, whereas both HGC and
CGC\_ES yield smooth distributions without such oscillations. The results
obtained with the two stability-preserving formulations are also consistent
with the reference data reported in
\cite{jagad2020investigation,jagad2021primitive}.

Figures~\ref{fig:compare_CGC_HGC}(e) and
\ref{fig:compare_CGC_HGC}(f) show the corresponding time histories of the
drag coefficient $C_d$ and lift coefficient $C_l$. Panel (e) shows the CGC
results, which exhibit large non-physical oscillations associated with
transient amplification of numerical perturbations. Panel (f) shows the HGC
and CGC\_ES results; both formulations produce bounded, regular force
histories characteristic of periodic vortex shedding and yield a Strouhal
number of $\mathit{St}=0.17$, in agreement with
\cite{MittalDBNVL:2008}. Thus, both CGC\_ES and HGC demonstrate the benefit of
satisfying the reconstruction-stability criterion, while HGC additionally
retains the compact nearest-neighbour stencil.

\begin{table}
 \caption{ Quantitative comparison of the flow past a circular cylinder with the literature. $C_{d,f}, C_{d,p}$ are the friction and pressure components of the drag coefficient $C_d$, respectively. $C_{pb}$ is the base (rare stagnation-point) pressure coefficient. {\it St} is the Strouhal frequency.}
 \centering
\resizebox{10cm}{!}{
\begin{tabular}[t]{|c|cccc|cccc|}
\hline
  Re 		& \multicolumn{4}{c|}{40}&\multicolumn{4}{c|}{100}\\
\hline
 & Henderson & Mittal & HGC	& CGC\_ES		& Henderson & Mittal & HGC	& CGC\_ES	\\
\hline
$C_{d,f}$	& 0.51	& -- 	& 0.52 & 0.53 	& 0.34	& --	& 0.33	& 0.33	\\
$C_{d,p}$	& 1.02	& --	& 1.07	& 1.02	& 1.00	& --	& 1.05	& 1.00	\\
$C_{d}$	& 1.53	& 1.53	& 1.58 & 1.54 & 1.34	& 1.35	& 1.37 & 1.34\\
$-C_{pb}$	& 0.52	& 0.52	& 0.50 & 0.50 & 0.73	& 0.74	& 0.74 & 0.73\\
{\it St}		& --	& --	& --	& --	& 0.17	& 0.17	& 0.17 & 0.17 \\
\hline
\end{tabular}
}
\label{table:cylinder}
\end{table}
Simulations were also performed at
$Re=1,20,40,60,80,95,105,$ and $150$. CGC produced spurious oscillations
throughout this range. The CGC simulations at $Re<100$ eventually diverged,
whereas those at $Re\geq100$ did not diverge over the simulated time interval
but developed non-physical oscillations with pronounced transient growth. In
contrast, HGC produced stable solutions for all Reynolds numbers considered.
Selected statistics evaluated after the initial transient are reported in
Table~\ref{table:cylinder} for $Re=40$ and $100$. The drag coefficient,
base-pressure coefficient, and Strouhal number obtained with HGC and CGC\_ES
are in good agreement with the spectral-element results of
\cite{Henderson:1995} and the immersed-boundary results of
\cite{MittalDBNVL:2008}.b
\subsection{Flow past an airfoil}\label{sec:airfoil}
\begin{figure}[htp]
 \centering
 \begin{tabular}{cc}
   \begin{minipage}{0.45\linewidth}
     \centering
     \includegraphics[width=2.75in,trim={0.8in 0.35in 0.8in 0.35in},clip]{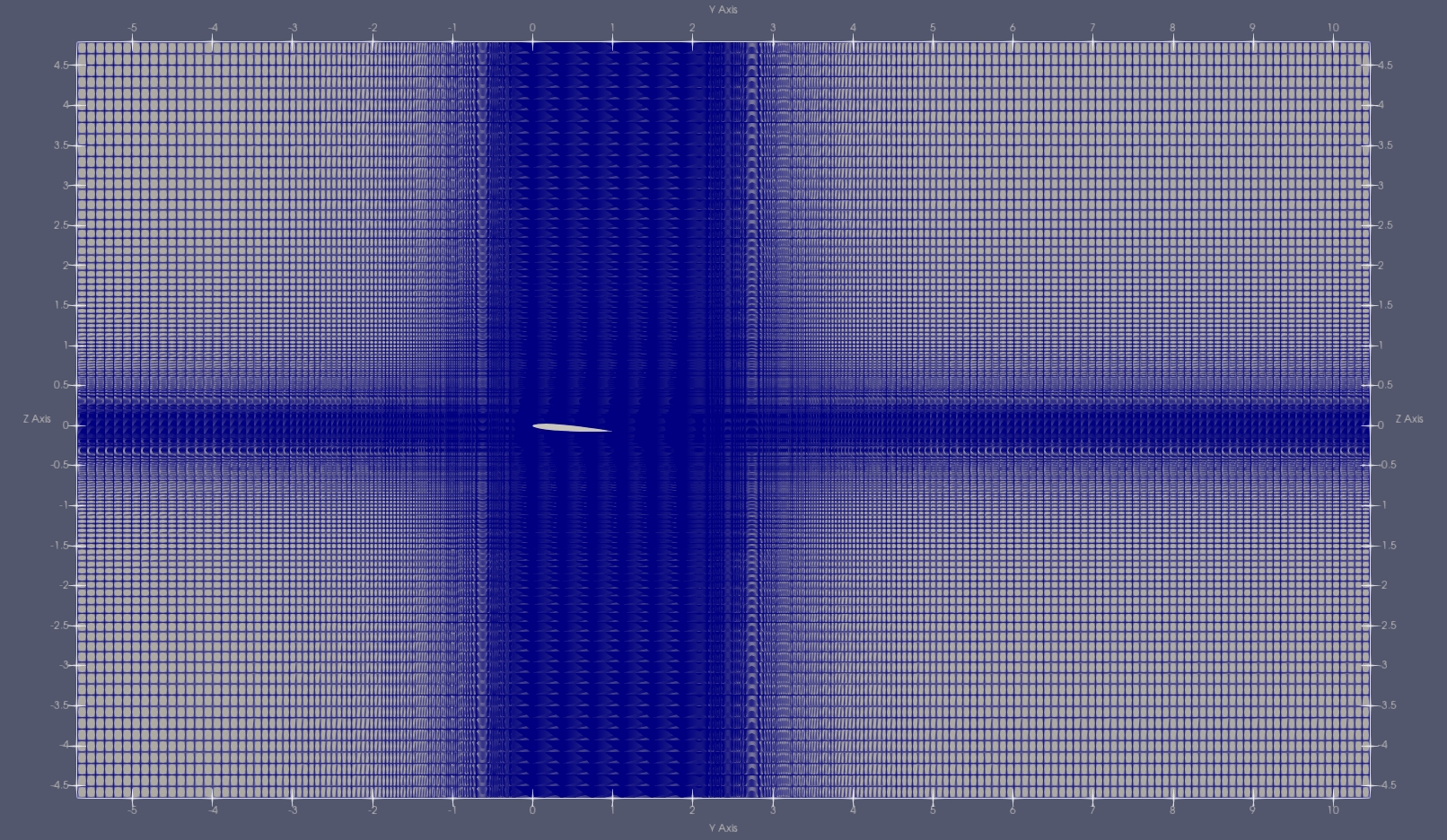}\\
     {\scriptsize(a) Computational domain and mesh}
   \end{minipage}
   \begin{minipage}{0.45\linewidth}
     \centering
     \includegraphics[width=2.75in,trim={0.in 0.in 0.in 0.in}]{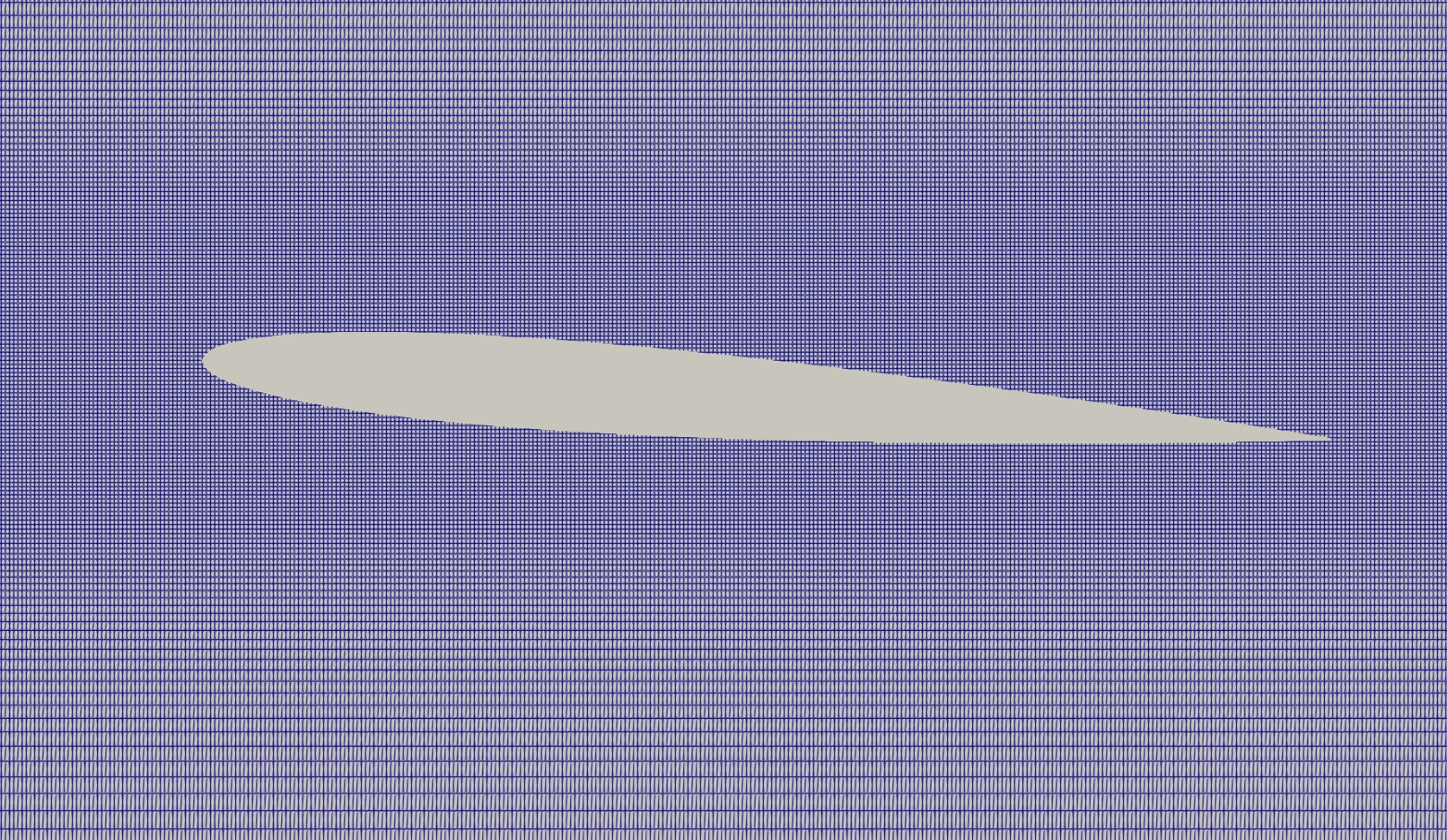}\\
     {\scriptsize(b) Mesh distribution near airfoil}
   \end{minipage}
 \end{tabular}
\caption{
Computational domain and Cartesian mesh used for the airfoil simulation at
$Re=2000$: (a) computational domain and mesh; (b) detailed mesh distribution
near the airfoil.
}
 \label{fig:mesh_airfoil}
\end{figure}

Incompressible flow past an airfoil of chord length $l$ is simulated at
$Re=2000$ and an angle of attack of $4^\circ$ using the CGC and HGC reconstruction strategies. The computational domain has dimensions $16l\times9l\times0.4l$ and is discretized using a
$512\times256\times4$ Cartesian grid. As shown in Fig.~\ref{fig:mesh_airfoil}, the mesh is locally uniform in the vicinity of the airfoil, with a spacing of $0.0036l$, and is gradually stretched toward
the outer boundaries with a stretching ratio of $3\%$.


\begin{figure}[htp]
 \centering
 \begin{tabular}{cc}
   \begin{minipage}{0.49\linewidth}
     \centering
     \includegraphics[width=2.25in]{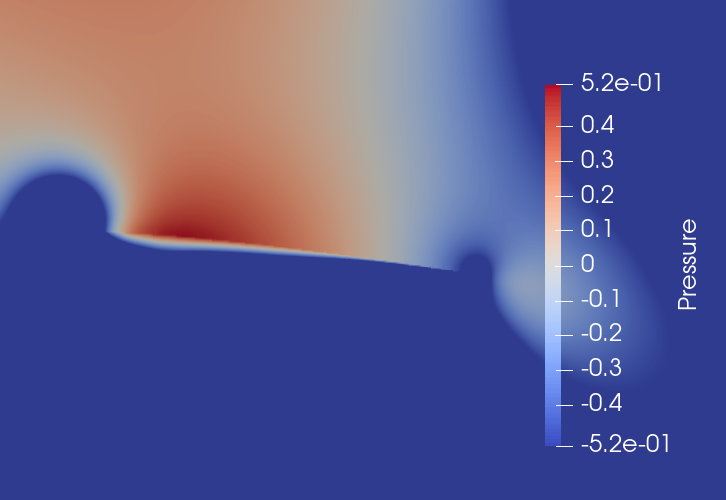}\\
     {\scriptsize(a) Pressure with CGC}
   \end{minipage}
   \begin{minipage}{0.49\linewidth}
     \centering
     \includegraphics[width=2.75in]{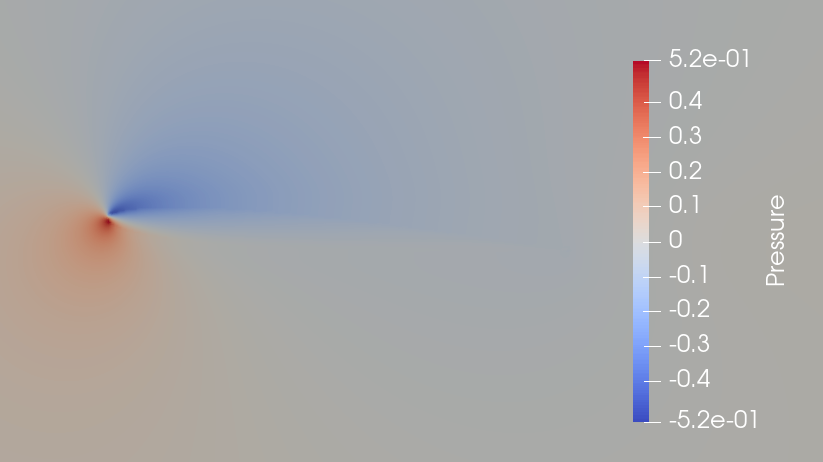}\\
     {\scriptsize(b) Pressure with HGC}
   \end{minipage}
   \\
   \begin{minipage}{0.49\linewidth}
     \centering
     \includegraphics[width=2.8in]{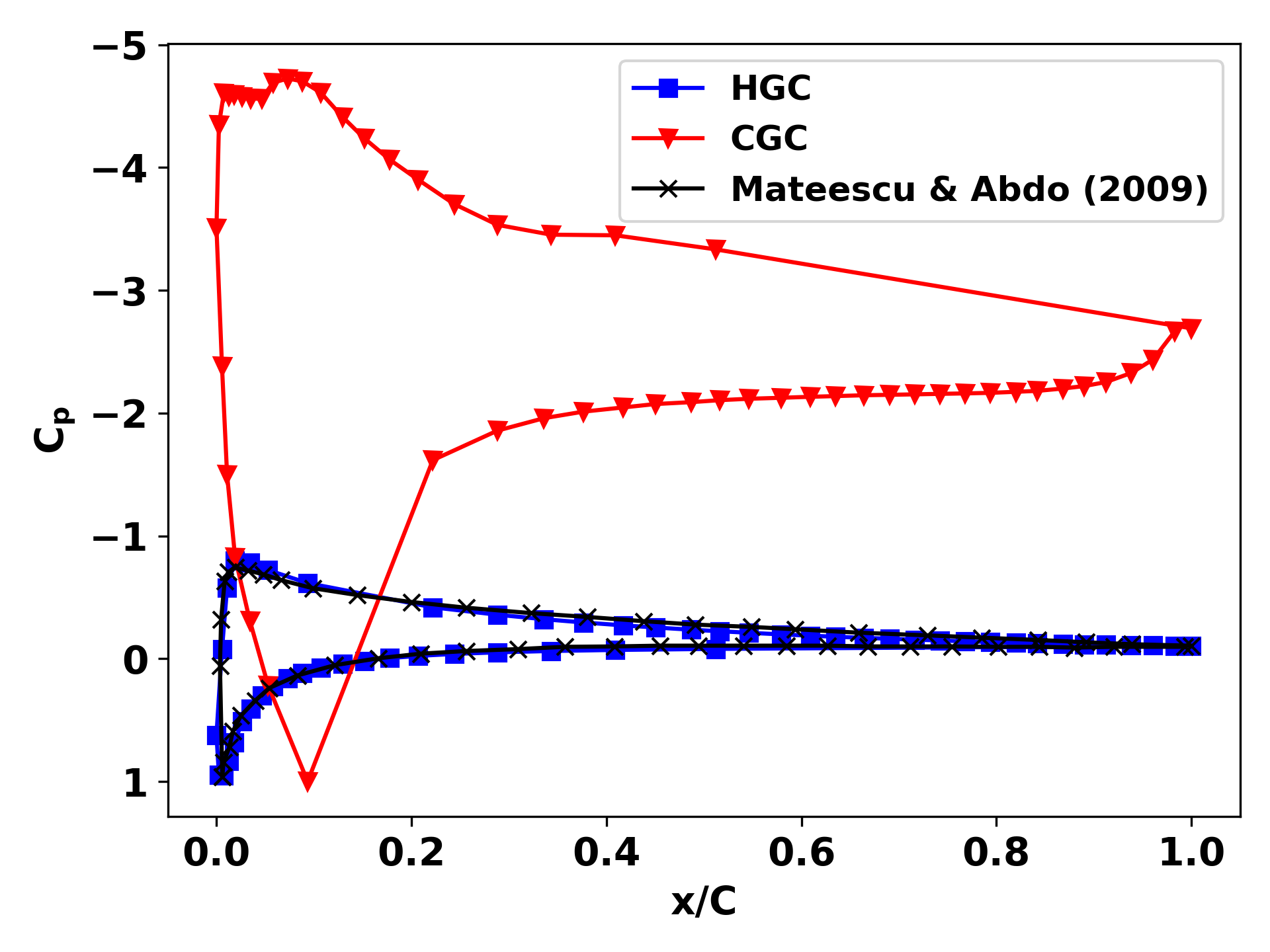}\\
     {\scriptsize(c) Distribution of $C_p$}
   \end{minipage}
   \begin{minipage}{0.49\linewidth}
     \centering
     \includegraphics[width=2.8in]{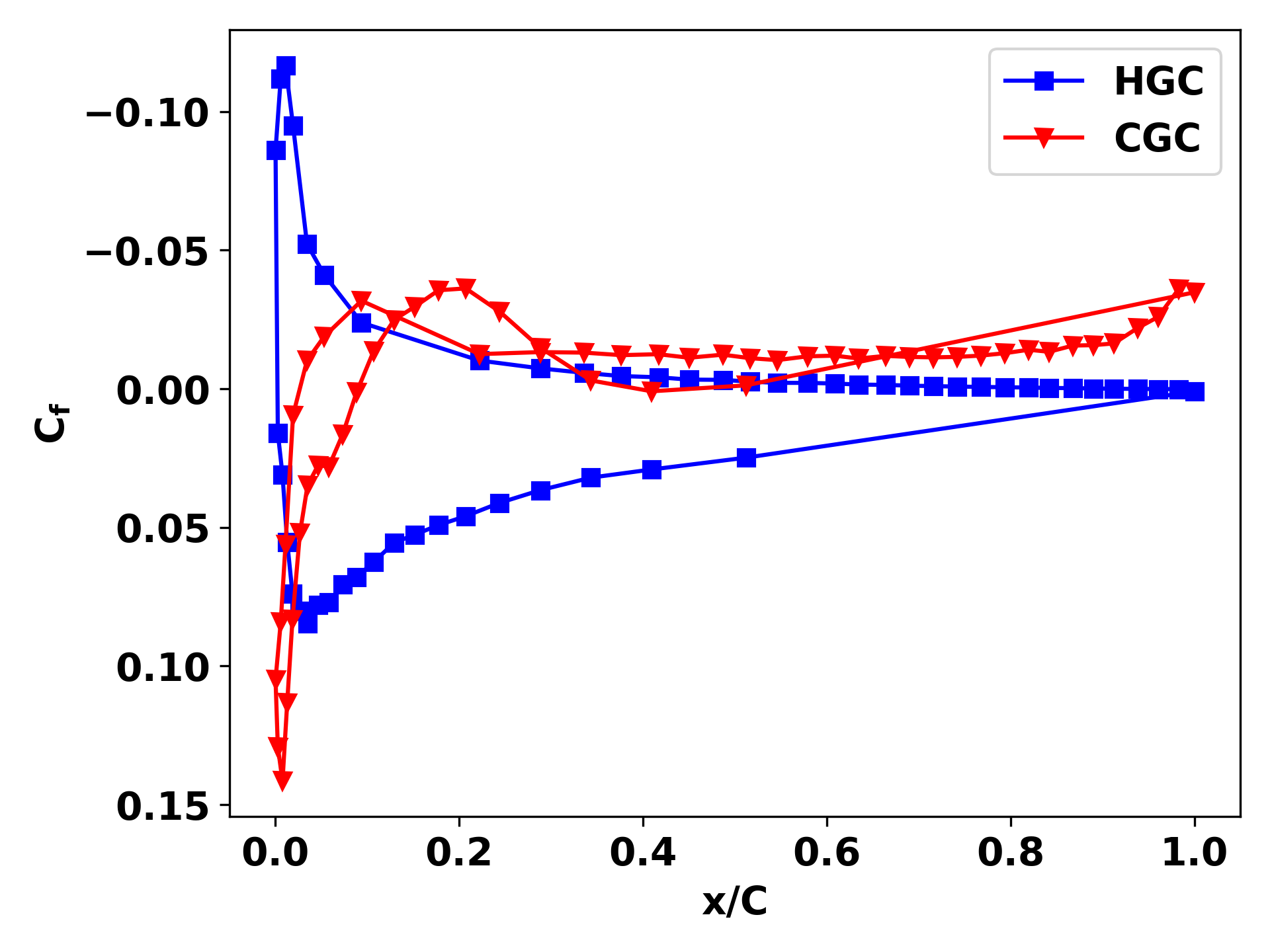}\\
     {\scriptsize(d) Distribution of $C_f$}
   \end{minipage}
   \\
   \begin{minipage}{0.49\linewidth}
     \centering
     \includegraphics[width=2.8in]{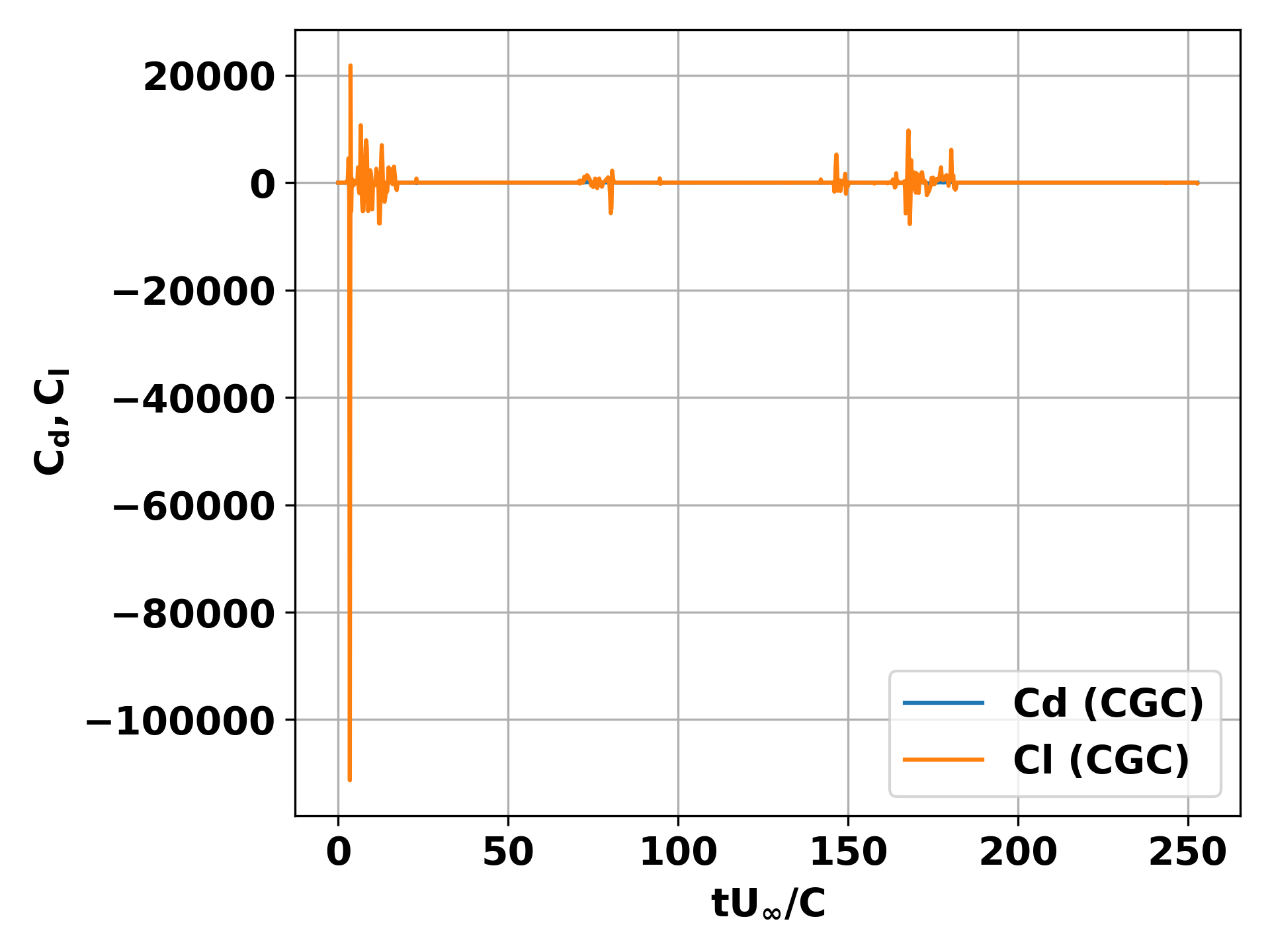}\\
     {\scriptsize(e) Time evolution of $C_d, C_l$ (CGC)}
   \end{minipage}
   \begin{minipage}{0.49\linewidth}
     \centering
     \includegraphics[width=2.8in]{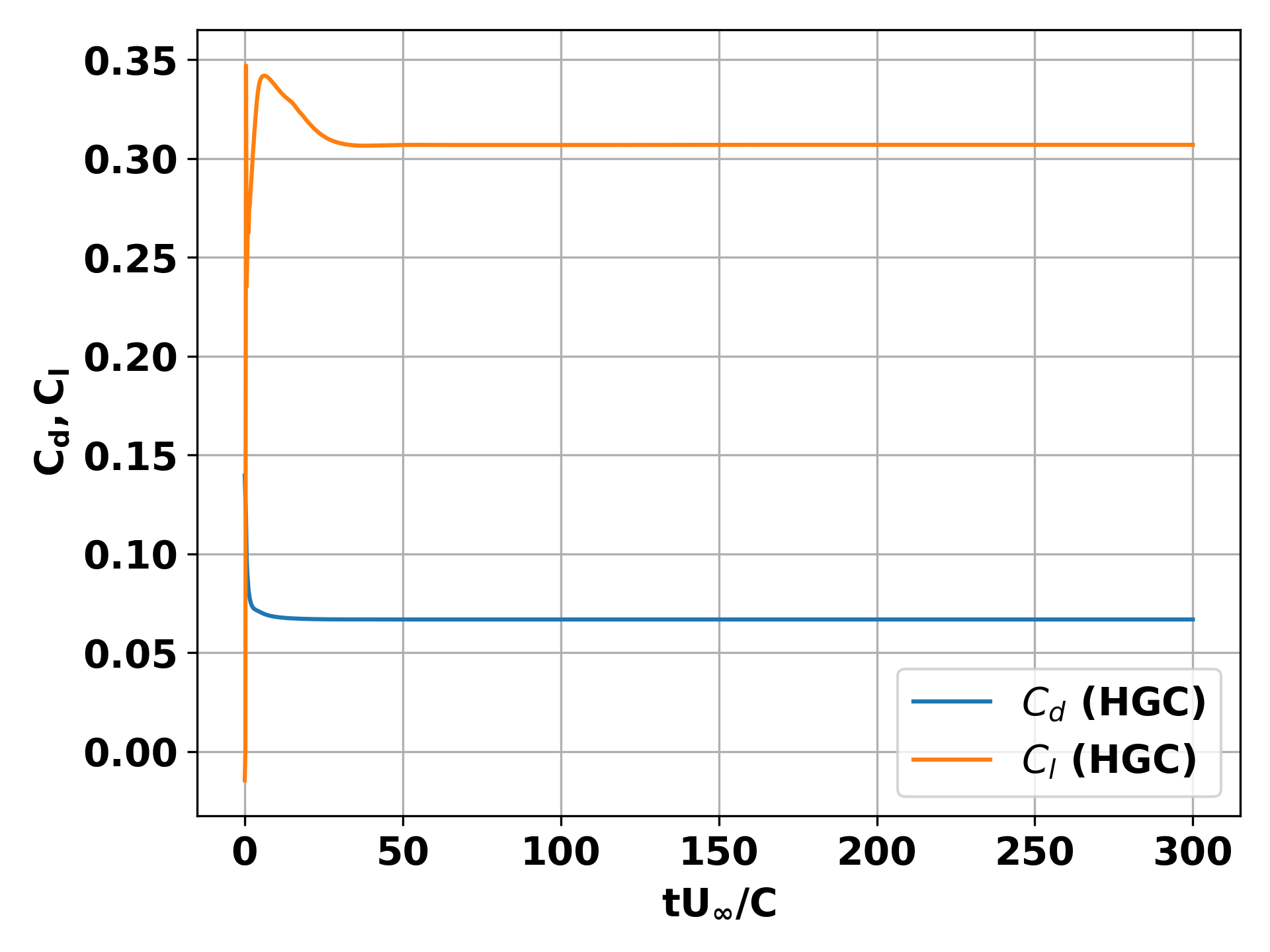}\\
     {\scriptsize(f) Time evolution of $C_d, C_l$ (HGC)}
   \end{minipage}
 \end{tabular}
\caption{
Flow past an airfoil at $Re=2000$ and an angle of attack of $4^\circ$.
(a) Pressure field at $t=123.8\,l/U_\infty$ obtained with CGC;
(b) pressure field at the same instant obtained with HGC;
(c) time-averaged pressure coefficient $C_p$;
(d) time-averaged skin-friction coefficient $C_f$;
(e) time histories of the drag and lift coefficients obtained with CGC;
(f) time histories of the drag and lift coefficients obtained with HGC.
The reference pressure-coefficient data of Mateescu and
Abdo~\cite{mateescu2010analysis} are included in panel (c).
Here, $l$ denotes the airfoil chord length.
}
 \label{fig:compare_CGC_HGC_airfoil}
\end{figure}

Figures~\ref{fig:compare_CGC_HGC_airfoil}(a) and
\ref{fig:compare_CGC_HGC_airfoil}(b) compare the pressure fields obtained
with CGC and HGC, respectively. HGC produces a smooth pressure distribution
around the airfoil, whereas CGC exhibits large non-physical pressure
excursions and spurious gradients, particularly near the leading edge and
along the suction side.

The corresponding time-averaged pressure and skin-friction distributions are
shown in Figs.~\ref{fig:compare_CGC_HGC_airfoil}(c) and
\ref{fig:compare_CGC_HGC_airfoil}(d). The pressure coefficient obtained with
HGC agrees closely with the reference distribution of Mateescu and
Abdo~\cite{mateescu2010analysis}, and the corresponding skin-friction
distribution remains smooth. In contrast, CGC produces substantial deviations
in both quantities, including incorrect pressure loading on the suction side
and non-physical extrema in the skin-friction coefficient near the leading
edge. The surface gradients used to evaluate $C_f$ and the aerodynamic forces
are computed using the analytical embedded-boundary expressions presented in
Sec.~\ref{sec:surface_quantities}; no separate surface-reconstruction or
filtering procedure is applied.

Figures~\ref{fig:compare_CGC_HGC_airfoil}(e) and
\ref{fig:compare_CGC_HGC_airfoil}(f) show the time histories of the drag and
lift coefficients. The CGC results exhibit intermittent large-amplitude
fluctuations, indicating transient amplification of numerical perturbations
generated near the embedded boundary. In contrast, the HGC force coefficients
approach $C_d=0.067$ and $C_l=0.307$ after the initial transient. These values
are in reasonable agreement with the values $C_d=0.0806$ and $C_l=0.27$
reported in \cite{ukken2019aerodynamic}. Although temporal averaging attenuates
the CGC fluctuations in the mean $C_p$ and $C_f$ distributions, the
instantaneous force histories clearly reveal the underlying numerical
instability. Overall, the airfoil results are consistent with the stability
behavior observed in the cylinder case and demonstrate the greater robustness
of the HGC reconstruction for this configuration.

\section{Conclusions}\label{sec:conclusions}
A direct analytical and non-iterative ghost-cell reconstruction framework has
been presented for Cartesian-grid embedded-boundary methods. Analytical
bilinear and trilinear reconstruction expressions were derived for Dirichlet
and Neumann boundary conditions in two and three dimensions. These expressions
impose the boundary condition directly at the embedded boundary, eliminating
the intermediate image-point reconstruction used in conventional approaches.
The resulting formulation avoids matrix inversion, precomputed interpolation
weights, and storage of reconstruction coefficients.

For the Cartesian reconstruction stencil considered here, the dependencies
among neighboring ghost cells form a directed acyclic graph. The resulting
topological ordering partitions the ghost cells into dependency levels
$L_1,\ldots,L_n$, allowing each ghost cell to be reconstructed once all values
required by its stencil are available. This eliminates iterative ghost-cell
updates and provides a level-by-level alternative to repeated reconstruction
sweeps and their associated communication stages in distributed-memory
implementations.

The respective roles of reconstruction stability and stencil compactness were
examined using three ghost-cell configurations. CGC retains classical
ghost-cell placement and a compact nearest-neighbour stencil, but does not
necessarily satisfy the reconstruction-stability criterion. CGC\_ES retains
the same ghost-cell placement while extending the stencil wherever required
to satisfy the criterion. HGC instead modifies the admissible ghost-cell
placement, allowing the same stability requirement to be satisfied while
retaining the compact nearest-neighbour Cartesian stencil.

The cylinder simulations provide a direct assessment of the stability
criterion. CGC exceeds the prescribed stability limit and develops spurious
oscillations in the pressure, surface coefficients, and force histories. In
contrast, both CGC\_ES and HGC satisfy the criterion and produce stable
solutions in agreement with reference results. These comparisons indicate
that the stability criterion previously obtained from linear analysis remains
a useful indicator of reconstruction stability for the nonlinear
incompressible Navier--Stokes cases considered here. They also demonstrate
the additional advantage of HGC: stability is achieved without enlarging the
reconstruction stencil. The airfoil simulations further corroborate the
improved robustness of HGC, which produces physically consistent pressure and
skin-friction distributions together with stable aerodynamic-force histories,
whereas CGC exhibits non-physical near-boundary oscillations.

The same analytical reconstruction relations also provide solution values and
Cartesian gradients directly on the embedded boundary. Pressure and viscous
forces, including drag and lift, can therefore be evaluated without a separate
surface-interpolation, reconstruction, or filtering procedure. Taken together,
the proposed framework addresses indirect image-point evaluation, iterative
ghost-cell coupling, reconstruction stability, and stencil compactness within
a unified formulation. It is therefore well suited to Cartesian-grid
embedded-boundary simulations in which direct boundary enforcement, compact
stencils, and non-iterative reconstruction are desired.

\section*{Acknowledgements}
This research was supported by the Emirates Nuclear Technology Center (KU-ENTC),
a collaboration among Khalifa University of Science and Technology, Emirates
Nuclear Energy Company, and the Federal Authority for Nuclear Regulation, and
by the Research Center for Mathematical \& Computational Intelligence (KU-MCI)
at Khalifa University of Science and Technology.
\appendix
\section{Proof of reconstruction formulae}\label{sec:proof}
\subsection{Two dimensions}\label{sec:proof2d}
\begin{figure}[t]
 \centering
 \begin{minipage}{0.45\linewidth}
 \centering
 \includegraphics[width=2.5in,trim={0.6cm 0.6cm 2cm 2.cm},clip]{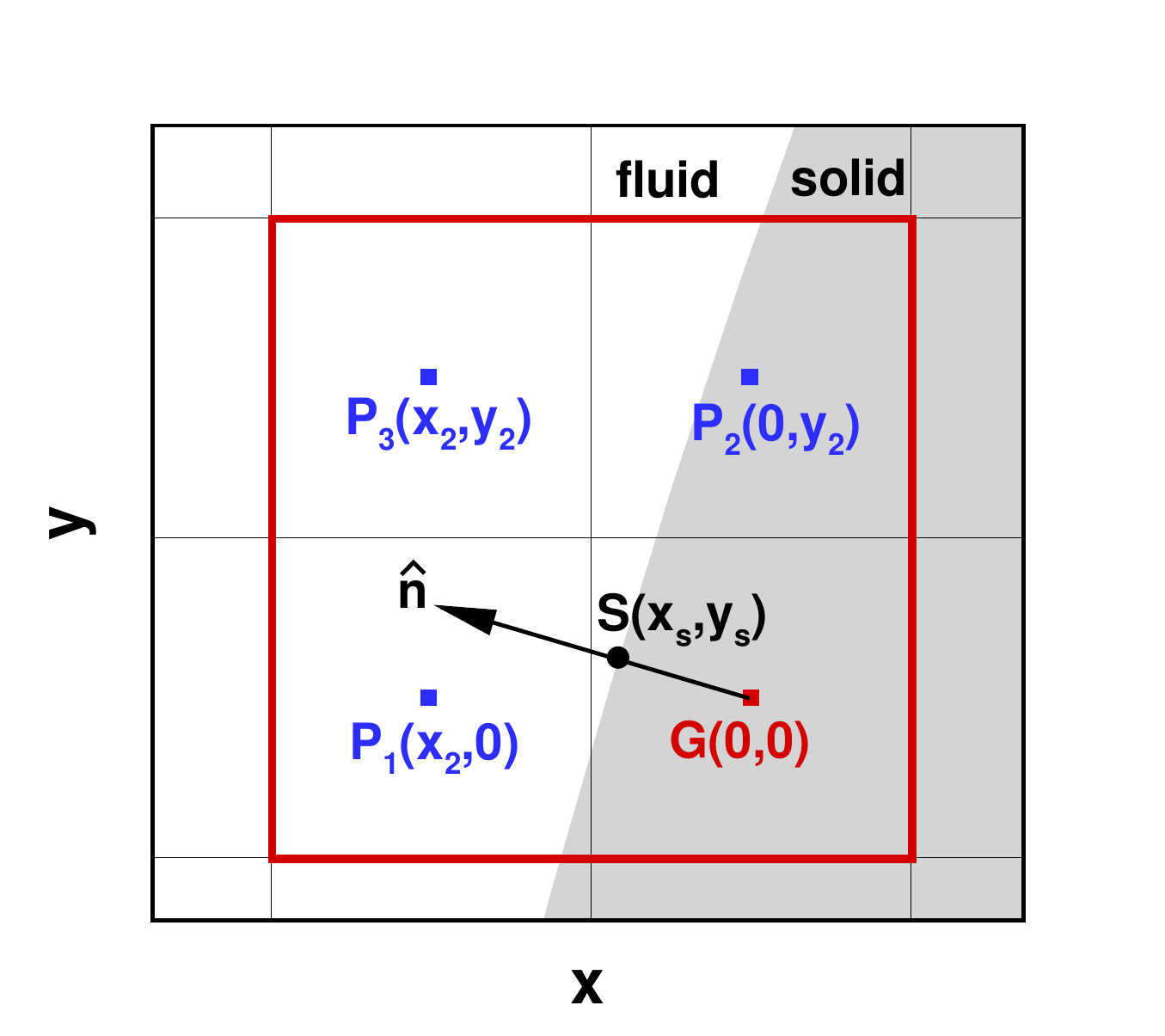} 
 \end{minipage}
 \caption{Stencil used in a bilinear interpolation for ghost cell $G$: $P_1, P_2, P_3$ are the neighbour cells, depending upon the orientation of the surface normal ($\hat{n}$), in a Cartesian grid and $S$ is the intersection of the embedded boundary with the surface normal ($\hat{n}$) passing through $G$. Here, a local coordinate system is used with origin at $G$.}
 \label{fig:stencil_2d_local}
\end{figure}
To simplify the derivation of reconstruction formulae for the ghost cell value, let us consider a local coordinate system with origin at the ghost cell $G$ so that $(x^{'},y^{'})=(x,y)-(x_1,y_1)$. Then the coordinates of all these points in the local coordinate system, by ignoring the superscript, can be written as, $G=(0,0), P_1=(x_2,0), P_2=(0,y_2), P_3=(x_2,y_2), S=(x_s,y_s)$ as shown in Fig. \ref{fig:stencil_2d_local}. 
\begin{eqnarray}
(x_2,y_2)&=& (\sgn{(n_x)}\Delta x, \sgn{(n_y)}\Delta y).
\end{eqnarray}
\subsubsection{Dirichlet BC}
In a local coordinate system with origin at $G(x_1,y_1)$, we have a linear system of equations,
\begin{equation}
\underbrace{\left\lbrace \begin{array}{c}\phi_1\\\phi_2\\\phi_3\\\phi_4\end{array}\right\rbrace}_{\{\phi\}}
= \underbrace{\left[ \begin{array}{cccc}
1	&  x_2 	&  0 		& 0	\\
1	&  0 		&  y_2 	& 0	\\
1	&  x_2 	&  y_2 	&  x_2 y_2\\
1	&  x_s 	&  y_s & x_s y_s
\end{array}\right]}_{[V]}
\underbrace{\left\lbrace \begin{array}{c} C_0\\C_1\\C_2\\C_3\end{array}\right\rbrace }_{\{C\}}
\end{equation}
where $\{\phi_i\}, i=1,2,3$ are the values evaluated at $P_1, P_2, P_3$, respectively, and $\phi_4=\phi_s$ evaluated at $S$. 
The unknown coefficients can be evaulated using,
\begin{equation}
 \{C\} = [V]^{-1} \{\phi\}.
\end{equation}
We, then, have the ghost cell value
\begin{equation}
 \phi_{G}=\phi(0,0) = C_0 = \left. \sum_{j=1}^{4} V^{-1}_{1j} \phi_j \right.
\end{equation}
Note that we need to compute only the first row of $V^{-1}$ to obtain the ghost cell value, $\phi_G$, which is given by, $V^{-1}_{1j} = \left\{ \frac{-x_s}{x_2 - x_s}, \frac{-y_s}{y_2 - y_s }, \frac{-x_s y_s}{(x_2 - x_s)(y_2 - y_s )},\frac{x_2 y_2}{(x_2 - x_s)(y_2 - y_s )} \right\}$.
Therefore, $\phi_G$, in the local coordinate system, can be expressed as
\begin{eqnarray*}
{\phi_G = \omega_1\phi_1+ \omega_2\phi_2 -\omega_1\omega_2\phi_3 +(1-\omega_1)(1-\omega_2) \phi_s},\quad
 \omega_1 = \frac{-x_s}{x_2 - x_s},\quad
 \omega_2 =\frac{-y_s} {y_2 - y_s} 
\end{eqnarray*}
Replace $(x, y)$ with $(x-x_1, y-y_1)$ in the above equation to express $\phi_G$ in the global coordinate sytem which yields Eq. \eqref{eqn:phi_G_Dir_2d}.
\subsubsection{Neumann BC}
In a local coordinate system with origin at $G(x_1,y_1)$, we have a linear system of equations,
\begin{equation}
\underbrace{\left\lbrace \begin{array}{c}\phi_1\\\phi_2\\\phi_3\\\phi_4\end{array}\right\rbrace}_{\{\phi\}}
= \underbrace{\left[ \begin{array}{cccc}
1	&  x_2 	&  0 		& 0	\\
1	&  0 		&  y_2 	& 0	\\
1	&  x_2 	&  y_2 	&  x_2 y_2\\
0	&  n_x 	&  n_y 	& n_x y_s + n_y x_s
\end{array}\right]}_{[V]}
\underbrace{\left\lbrace \begin{array}{c} C_0\\C_1\\C_2\\C_3\end{array}\right\rbrace }_{\{C\}}
\end{equation}
where $\{\phi_i\}, i=1,2,3$ are the values evaluated at $P_1,P_2,P_3$, respectively, and $\phi_4=\bdelux{\phi}{n}$ evaluated at $S$.
The ghost cell value is $
 \phi_{G}=\phi(0,0) = C_0 = \left. \sum_{j=1}^{4} V^{-1}_{1j} \phi_j \right.$ which can be expressed in the global coordinate system as,
\begin{eqnarray}
{\phi_G = \omega_1\phi_1+ \omega_2\phi_2 +(1-\omega_1-\omega_2)\phi_3 +\omega_s \bdelux{\phi}{n}.}
\end{eqnarray}
where,
\begin{eqnarray*}
 \omega_1 &=&   \frac{(a_{2} + b_{1}  )}{(a_{2}+b_{2})}, \quad\quad
 \omega_2 =   \frac{(a_{1} + b_{2}  )}{(a_{2}+b_{2})}, \quad\quad
 \omega_s = -\frac{(x_2-x_1)(y_2-y_1)}{(a_{2}+b_{2})},\\
 a_{1} &=& n_x(y_1-y_s), \quad\quad
 a_{2} = n_x(y_2-y_s), \quad\quad
 b_{1} = n_y(x_1-x_s), \quad\quad
 b_{2} = n_y(x_2-x_s).	
 \end{eqnarray*}
Note that if $n_x=0$, then $x_s=x_1$ and if $n_y=0$, then $y_s=y_1$.
Replace $(x, y)$ with $(x-x_1, y-y_1)$ in the above equation to express $\phi_G$ in the global coordinate sytem which yields Eq. \eqref{eqn:phi_G_Neu_2d}.

\subsection{Three dimensions}\label{sec:proof3d}
If ($i,j,k$) are the Cartesian indices of the ghost cell $G$, then the indices of the diagonally opposite cell ($P_7$) in the stencil, relative to $G$, can be written as,
\begin{eqnarray}
(i_7,j_7,k_7)&=& (i+e_x, j+e_y, k+e_z).
\end{eqnarray}
Note that, since $x_2\neq x_1, y_2\neq y_1, z_2\neq z_1$ for the trilinear interpolation, all the components of $e$ must be strictly either +1 or -1. However, in a computer program, if any component of $e$ is 0, e.g., $n_x=0$ gives us $e_x=0$, it can be arbitrarily set to either $+1$ or -1 (+1 in our case) which does not affect the reconstruction because the weightage coefficients for the corresponding nodes become zero. 

For simplicity, let us consider a local coordinate system with origin at the ghost cell center $G$ so that $(x^{'},y^{'},z^{'})=(x,y,z)-(x_1,y_1,z_1)$. Then the coordinates of all these points in the local coordinate system, by ignoring the superscript, can be written as, 
\begin{eqnarray*}
P_1&=&(x_2,  0,  0), \hspace{1cm} ~~P_2=(0,y_2,0), \hspace{1cm} ~P_3=(0,0,z_2), \\
P_4&=&(x_2,y_2,0), \hspace{1cm} ~P_5=(0,y_2,z_2), \hspace{1cm} P_6=(x_2,0,z_2), \\
P_7&=&(x_2,y_2,z_2), \hspace{1cm} S=(x_s,y_s,z_s),  \hspace{1cm} ~G=(0,0,0).
\end{eqnarray*}
and
\begin{eqnarray}
(x_2,y_2,z_2)&=& (e_x\Delta x, e_y\Delta y, e_z\Delta z).
\end{eqnarray}

\subsubsection{Dirichlet BC}
In a local coordinate system with origin at $G(x_1,y_1,z_1)$, we have a linear system of equations,
\begin{equation}
\underbrace{\left\lbrace \begin{array}{c}\phi_1\\\phi_2\\\phi_3\\\phi_4\\\phi_5\\\phi_6\\\phi_7\\\phi_8\end{array}\right\rbrace}_{\{\phi\}}
= \underbrace{\left[ \begin{array}{cccccccc}
1	&  x_2 	&  0 		& 0		& 0 	& 0 & 0	& 0\\
1	&  0 		&  y_2 	& 0		& 0 	& 0 & 0	& 0\\
1	&  0 		&  0	 	& z_2	& 0 	& 0 & 0	& 0\\
1	&  x_2 	&  y_2 	& 0		&  x_2 y_2 & 0 & 0	& 0\\
1	&  0	 	&  y_2 	& z_2	&  0 & y_2 z_2 & 0	& 0\\
1	&  x_2 	&  0	 	& z_2	&  0 & 0 & x_2 z_2	& 0\\
1	&  x_2 	&  y_2 & z_2	&  x_2 y_2 & y_2 z_2 & x_2 z_2	& x_2 y_2 z_2\\
1	&  x_s 	&  y_s & z_s	&  x_s y_s & y_s z_s & x_s z_s	& x_s y_s z_s
\end{array}\right]}_{[V]}
\underbrace{\left\lbrace \begin{array}{c} C_0\\C_1\\C_2\\C_3\\C_4\\C_5\\C_6\\C_7\end{array}\right\rbrace }_{\{C\}}
\end{equation}
where $\{\phi_i\}, i=1,2,...,7$ are the values evaluated at $P_1,P_2, ...,P_7$, respectively, and $\phi_8=\phi_s$ evaluated at $S$. 
The unknown coefficients can be evaulated using,
\begin{equation}
 \{C\} = [V]^{-1} \{\phi\}\label{eqn:trilinear}
\end{equation}
Since the ghost cell value $ \phi_{G}=\phi(0,0,0) = C_0 = \left. \sum_{j=1}^{8} V^{-1}_{1j} \phi_j \right.$, we need to compute only the first row of $V^{-1}$ which is given by,
$
V^{-1}_{1j} = \left\{ \frac{-x_s}{X}, \frac{-y_s}{Y}, \frac{-z_s}{Z}, \frac{-x_s y_s}{XY}, \frac{-y_s z_s}{YZ}, \frac{-x_s z_s}{XZ}, \frac{-x_s y_s z_s}{XYZ}, \frac{x_2 y_2 z_2}{XYZ} \right\}
$
where, $X = x_2 - x_s, Y = y_2 - y_s, Z = z_2 - z_s.$
In the local coordinate system $\phi_G$ can be expressed as,
\begin{eqnarray}
{\phi_G = \omega_1\phi_1+ \omega_2\phi_2+ \omega_3\phi_3 -\omega_1\omega_2\phi_4 -\omega_2\omega_3\phi_5 -\omega_1\omega_3\phi_6+\omega_1\omega_2\omega_3\phi_7+(1-\omega_1)(1-\omega_2)(1-\omega_3) \phi_s}
\end{eqnarray}
where,
\begin{eqnarray*}
 \omega_1 = \frac{-x_s}{x_2 - x_s}, \quad
 \omega_2 =\frac{-y_s} {y_2 - y_s}, \quad
 \omega_3 = \frac{-z_s}{z_2 - z_s}.
\end{eqnarray*}
Replace $(x, y, z)$ with $(x-x_1, y-y_1, z-z_1)$ in the above equation to express $\phi_G$ in the global coordinate sytem which yields Eq. \eqref{eqn:phi_G_Dir_3d}.

\subsubsection{Neumann BC}
In a local coordinate system with origin at $G(x_1,y_1,z_1)$, we have a linear system of equations,
\begin{equation}
\underbrace{\left\lbrace \begin{array}{c}\phi_1\\\phi_2\\\phi_3\\\phi_4\\\phi_5\\\phi_6\\\phi_7\\\phi_8\end{array}\right\rbrace}_{\{\phi\}}
= \underbrace{\left[ \begin{array}{llllllll}
1	&  x_2 	&  0 		& 0		& 0 	& 0 & 0	& 0\\
1	&  0 		&  y_2 	& 0		& 0 	& 0 & 0	& 0\\
1	&  0 		&  0	 	& z_2	& 0 	& 0 & 0	& 0\\
1	&  x_2 	&  y_2 	& 0		&  x_2 y_2 & 0 & 0	& 0\\
1	&  0	 	&  y_2 	& z_2	&  0 & y_2 z_2 & 0	& 0\\
1	&  x_2 	&  0	 	& z_2	&  0 & 0 & x_2 z_2	& 0\\
1	&  x_2 	&  y_2 & z_2	&  x_2 y_2 & y_2 z_2 & x_2 z_2	& x_2 y_2 z_2\\
0	&  n_x 	&  n_y & n_z	&  x_s n_y+& y_sn_z+ & x_sn_z+& x_sy_sn_z+\\
	&   	&   & 	&  y_s n_x & z_sn_y & z_sn_x	& y_sz_sn_x+\\
	&   	&   & 	&   &  & 	& x_sz_sn_y
\end{array}\right]}_{[V]}
\underbrace{\left\lbrace \begin{array}{c} C_0\\C_1\\C_2\\C_3\\C_4\\C_5\\C_6\\C_7\end{array}\right\rbrace }_{\{C\}}
\end{equation}
where $\{\phi_i\}, i=1,2,...,7$ are the values evaluated at $P_1,P_2, ...,P_7$, respectively, and $\phi_8=\bdelux{\phi}{n}$ evaluated at $S$.
Since the ghost-cell value is
\[
\phi_G=\phi(0,0,0)=C_0=\sum_{j=1}^{8}V^{-1}_{1j}\phi_j,
\]
only the first row of $[V]^{-1}$ is required to obtain the reconstruction in
the local coordinate system. Transforming back to the global coordinate
system by replacing $(x,y,z)$ with $(x-x_1,y-y_1,z-z_1)$ yields the Neumann
reconstruction given in Eq.~\eqref{eqn:phi_G_Neu_3d}.
\bibliographystyle{elsarticle-num}
\bibliography{Misc/ibm_References,Misc/References}






\end{document}